\documentclass[aps,pre,superscriptaddress,twocolumn,floatfix,10pt,floatfix]{revtex4-2}
\usepackage{dcolumn,physics}
\usepackage{CJK}
\usepackage{graphicx,amsmath,amssymb}
\usepackage{comment,multirow,xcolor}
\usepackage{eucal,mathrsfs,dsfont,bm}
\usepackage{comment}

\usepackage[makeroom]{cancel}
\usepackage{color}
\definecolor{myred}{RGB}{250, 40, 90}
\definecolor{myblue}{RGB}{50, 130, 250}
\definecolor{mygreen}{RGB}{10, 140, 90}
\begin{document}
%-----------------------------------
\begin{CJK*}{UTF8}{}
\title{
Interplay between intraspecific suppression and environment in shaping biodiversity
}
%-----------------------------------
\author{Seong-Gyu Yang \CJKfamily{mj}{(양성규)}}
\affiliation{School of Computational Sciences, Korea Institute for Advanced Study, Seoul, 02455, Republic of Korea}
\affiliation{Asia Pacific Center for Theoretical Physics, Pohang, 37673, Republic of Korea}
\author{Hye Jin Park \CJKfamily{mj}{(박혜진)}}
\email[Corresponding author:]{hyejin.park@inha.ac.kr}
\affiliation{Department of Physics, Inha University, Incheon, 22212, Republic of Korea}
%-----------------------------------
\date{\today}
%-----------------------------------
\begin{abstract}
Understanding the mechanisms that sustain high biodiversity remains a central challenge.
MacArthur's classical consumer-resource model (MCRM) suggests that consumer diversity is limited by the number of available resources, yet empirical observations often exceed this bound.
To address this, we extend the generalized consumer-resource model by incorporating intraspecific suppression and analyze its effects using the dynamical mean-field theory.
Our results show that intraspecific suppression promotes biodiversity by preventing the emergence of dominant species and enabling more species to coexist, particularly in resource-rich environments.
Furthermore, our results provide analytical bounds on relative diversity, demonstrating that the number of coexisting consumer species can exceed the number of resource kinds.
This highlights the critical role of intraspecific suppression and environmental factors in promoting coexistence.
\end{abstract}
%-----------------------------------
\maketitle
\end{CJK*}
\date{\today}
%-----------------------------------
\section{Introduction}
\label{sec:intro}
By focusing on interactions between consumer species and resource kinds, the consumer-resource model~\cite{SF_MacArthur,MCRM_1, MCRM_2} provides powerful tools for understanding and predicting the quantitative dynamics of complex ecosystems.
A key application of this model lies in studying ecosystems with high biodiversity, where significant efforts have been dedicated to explaining how diverse consumer species coexist and finding what determines the number of coexisting species.
MacArthur's original consumer-resource model (MCRM) has shown that the number of coexisting consumer species cannot exceed the number of available resource kinds~\cite{NsNm1, NsNm2, NsNm3}.
Experimental evidence, such as studies on the coexistence of unicellular organisms~\cite{Gause_Experiment, Gause_Book} and observations on warbler species~\cite{Warbler_Habitat}, has consistently supported these theoretical predictions.

%High diversity in nature and explainable mechanisms
Although the MCRM offers a plausible upper limit on the number of coexisting consumer species, it is common to observe a greater diversity of consumer species than the number of available resource kinds in nature.
One such example is found in the ocean with phytoplankton~\cite{ParadoxPlankton1}.
Despite consuming a limited number of resources, phytoplankton communities display a remarkable diversity, with a greater number of species than the available resource kinds~\cite{Plankton_diversity1, Plankton_diversity2, Plankton_diversity3}.
To explain such high biodiversity, which is beyond the bound predicted by the MCRM, mechanisms such as temporal environmental fluctuations, spatial heterogeneity, and others have been proposed and provided successful explanations~\cite{ParadoxPlankton1,paradox_resolve,MCRM_Monod}.

%Introduction to the intraspecific suppression
A mechanism of growing interest~\cite{SF_Eco,SF_plankton,MCRM_Monod, Momeni_eLife2024} is intraspecific suppression that includes direct competition (e.g., mating conflicts) and growth suppression from pathogen transmission within species~\cite{SF_Chesson, MCRM_Monod} (e.g., the Kill the Winner hypothesis~\cite{KtW_model1, KtW_model2, KtW_model3, KtW_model4}).
It has been demonstrated that intraspecific suppression effectively promotes high biodiversity by preventing any single species from becoming overly dominant.
However, the effect of intraspecific suppression combined with environmental factors is not well-examined and understood.

%What we did in detail: Upper Bound of biodiversity and Effect of the environment
To address this gap, we employed the generalized MCRM (GCRM) with externally supplied resources.
A recent study~\cite{GCRM_Niche} has shown that even under abundant resource input, the GCRM predicts an upper bound on biodiversity---50\% of the resource pool for externally supplied resources, and at 100\% for self-renewing resources---in the absence of intraspecific suppression.
In contrast, our results demonstrate that this limit can be surpassed under specific ecological conditions.
In particular, we show that intraspecific suppression, when combined with sufficient external resource supply, allows the coexistence of more consumer species than resource types.

To understand this mechanism analytically, we applied the dynamical mean-field theory (DMFT) to handle disordered interactions~\cite{Cavity_Book1,Cavity_Book2,LV_Cavity,LV_Cavity2,LV_DMFT,LV_DMFT2,GCRM_Cavity,GCRM_DMFT,HMFT}.
This approach enabled us to derive theoretical bounds on relative diversity, defined as the ratio of surviving consumer species to available resource types. Our analysis reveals that resource availability modulates the effect of intraspecific suppression, thereby influencing relative diversity. These theoretical predictions are supported by numerical simulations, which show excellent agreement with our analytical results.

%----------------
%----------------
\section{Model}
\label{sec:model}
\begin{figure}[!t]
\centering\includegraphics[width=1.00\linewidth]{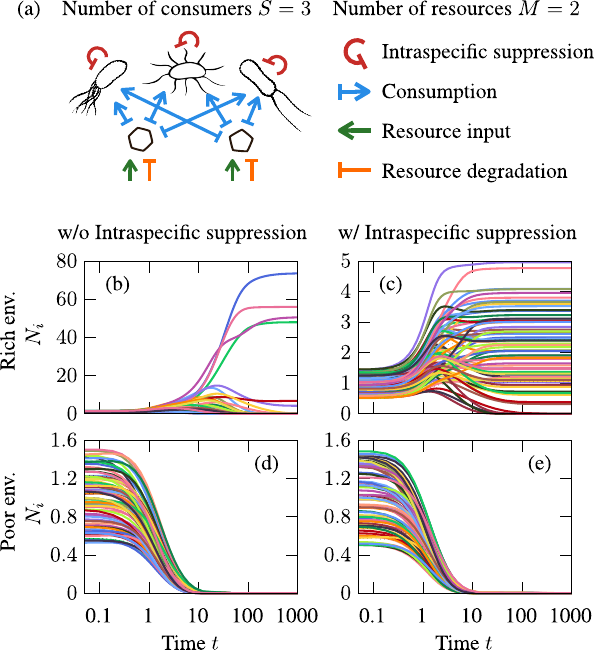}
\caption{
(a) Schematic of an ecological system with three consumer species ($S=3$) and two resource kinds ($M=2$).
Three bacteria species consume two externally supplied resources.
Red circular arrows indicate intraspecific suppression, blue straight arrows denote resource consumption, and green and orange arrows represent external input and degradation, respectively.
Growth at the sharp end ($\to$) is stimulated, while that at the blunt end ($\dashv$) is suppressed.
[(b)--(e)] Consumer abundance dynamics in ecosystems with $S=75$ and $M=50$, under resource-rich [(b) and (c)] and resource-poor [(d) and (e)] environments.
[(b) and (d)] show cases without intraspecific suppression, while [(c) and (e)] are with it.
In resource-rich environments, intraspecific suppression enhances biodiversity [(b) and (c)], whereas in resource-poor environments, it alone cannot sustain high diversity [(d) and (e)].
}
\label{fig:Schematic}
\end{figure}

The GCRM describes the abundance dynamics of ecological systems consisting of $S$ different consumer species and $M$ different kinds of resources~\cite{GCRM_Niche}.
By introducing the intraspecific suppression in the GCRM, we write the abundance dynamics of consumers and resources as
\begin{equation}
\begin{split}\label{eq:model_ext}
\dot{N}_i &=N_i \left( \epsilon_i \sum_\alpha C_{i\alpha} R_\alpha -m_i -h_i N_i \right),\\
\dot{R}_\alpha &=K_\alpha - \left( D_\alpha +  \sum_i N_i C_{i\alpha} \right)R_\alpha,
\end{split}
\end{equation}
where $N_i$ and $R_\alpha$ denote the abundance of consumer $i~( = 1, 2, \cdots, S )$ and resource $\alpha ~( = 1, 2, \cdots, M )$, respectively.
Consumer $i$ grows by taking resource $\alpha$ with a consumption rate $C_{i\alpha}$ and dies with a mortality rate $m_i$.
In the consumer dynamics, $\epsilon_i$ represents the trophic efficiency, capturing the conversion efficiency from consumed resources to consumer's biomass.
For simplicity, we set $\epsilon_i = \epsilon = 1$ for all consumers throughout this study.
The abiotic resource $\alpha$ is externally supplied with an input rate $K_\alpha$ and degraded with a degradation rate $D_\alpha$.
The last term in the consumer abundance dynamics indicates the intraspecific suppression, and $h_i$ is an intraspecific suppression coefficient of consumer $i$.
In the ecological literature, such mechanisms are often referred to as self-regulation.
In this work, we adopt the term intraspecific suppression to emphasize the explicit, quadratic form of negative feedback built into our model.
The schematic figure of the model with three consumer species ($S=3$) and two different resource kinds ($M=2$) is shown in Fig.~\ref{fig:Schematic}(a).
At the steady state, some consumers may go extinct, resulting in $S^*(\leq S)$ surviving consumers coexisting.
In contrast, resources never go extinct due to the external supply, thus the number of available resources remains constant at $M$, i.e., $M^*=M$.

To analyze ecological systems using statistical physics methods, we model a large ecological system where the number of consumer species $S$ and resource kinds $M$ are large $(S, M \gg 1)$, while maintaining a finite ratio $\gamma~(= M/S)$. Since we consider large systems, assigning all the values of $C_{i\alpha}$ is unfeasible.
In this situation, a random matrix approach provides a systematic way to describe the interactions in theoretical approaches~\cite{RMay,RandomMatrix,GCRM_Niche,GCRM_Cavity,DynamicPersistence,GCRM_DMFT,LV_Cavity,LV_Cavity2,LV_DMFT,LV_RandomMatrix,LV_RandomMatrix2,HMFT}.
The values of $C_{i\alpha}$ are randomly drawn from the Gaussian distribution $\mathcal{N}(\mu_C/M, \sigma^2_C/M)$ with the mean $\mu_C/M$ and variance $\sigma^2_C/M$.
Other parameters such as $m_i$, $K_\alpha$, and $D_\alpha$ are also independently drawn from Gaussian distributions, where the mean and standard deviation of a parameter $X$ are denoted by $\mu_X$ and $\sigma_X$, respectively.
To focus on investigating the effects of intraspecific suppression and environments, we vary $\mu_K$, and $\mu_h$, while keeping the following parameters fixed throughout this study: $\mu_C=1$, $\mu_m=1$, $\mu_D=1$, $\sigma_m=1/10$, $\sigma_m=1/10$, $\sigma_D=0$, and $\sigma_K=1/10$.
For the sake of simplicity, we set $h_i=h$ for all $i$, i.e., $\mu_h=h$ with $\sigma_h=0$.
A general study of cases with nonzero $\sigma_h$ is addressed in the Supplemental Material (SM)~\cite{SM}.

Through a brief investigation of consumer dynamics, we found that the interplay between intraspecific suppression $h$ and environmental condition $\mu_K$ is crucial to determining consumer species' diversity [see Figs.~\ref{fig:Schematic}(b)--(e)].
In the presence of intraspecific suppression, consumers are unable to attain large abundances, thereby leading to an increase in the number of small consumer communities and enhancing diversity [Figs.~\ref{fig:Schematic}(b) and (c)].
However, this effect diminishes in resource-poor environment, where species struggle to meet minimum energy requirements, which is shown in Figs.~\ref{fig:Schematic}(d) and (e).
Through theoretical derivation of the relative diversity $S^*/M^*$ at steady state and numerical integration of Eq.~\eqref{eq:model_ext}, we rigorously investigate these scenarios and present theoretical bounds of $S^*/M^*$.
A detailed explanation of the numerical methods is provided in SM~\cite{SM}, and all the codes are available in Ref.~[\onlinecite{my_git}].

%
%-----------------------------------
\section{Results}
\label{sec:results}
%--------------------
\subsection{Consumer and resource abundance distributions}
\label{ssec:Ps}
%--------------------
%

%

For externally supplied resources, the relative diversity $S^*/M^*$ is given by $\gamma^{-1}\phi_S$, where $\phi_S$ represents the surviving probability of consumers at steady state, i.e., $\phi_S=S^*/S$.
Since $\phi_S = \int_{+0}^{\infty}dN P(N)$, the behavior of $S^*/M^*$ can be understood through consumer abundance distribution $P(N)$, which requires knowledge of resource abundance distribution $P(R)$.
Therefore, to calculate $S^*/M^*$ at steady state, we obtain $P(N)$ and $P(R)$ first.

By applying DMFT~\cite{Cavity_Book1, Cavity_Bethe,Cavity_Book2,DMFT,DMFT_Book1,DMFT_Book2,DMFT1,DMFT2,HMFT} to Eq.~\eqref{eq:model_ext},
we obtain the effective mean-field dynamics (details are in SM~\cite{SM}).
Note that DMFT assumes all consumers follow the same effective dynamics, and all resources do as well, resulting in two effective equations for consumer abundance $N$ and resource abundance $R$.
The steady-state condition of the effective dynamics gives
\begin{equation}\label{eq:steady}
\begin{split}
0 =& N\left[ \epsilon \mu_C \langle R \rangle - \mu_m  - \left(h + \sigma_C^2  \langle \chi \rangle \right) N \right. \\
&\quad \left. + z_N \sqrt{\sigma^2_m + \epsilon^2\sigma_C^2 \langle R^2 \rangle } \right],\\
0 =& \mu_K - \left( \mu_D +\gamma^{-1}\mu_C \langle N \rangle \right) R - \gamma^{-1}\epsilon\sigma_C^2   \langle \nu \rangle R^2 \\
&\quad + z_R \sqrt{\sigma^2_K + \left(\sigma^2_D + \gamma^{-1}\sigma_C^2 \langle N^2 \rangle \right)R^2},
\end{split}
\end{equation}
where $z_N$ and $z_R$ are independent unit Gaussian random variables arising from the disorders in the model.
In Eq.~\eqref{eq:steady}, $\langle \cdot \rangle$ denotes the ensemble average.
The terms $\langle \nu \rangle$ and $\langle \chi \rangle$ represent the response functions of consumer and resource, respectively, where the consumer response function is given by $\langle \nu \rangle =- \langle \partial N / \partial \mu_m \rangle$, and the resource response function by $\langle \chi \rangle= - \langle \partial R / \partial \mu_D \rangle$.

Given that $P(z_X) = 1/\sqrt{2\pi} \exp(-z_X^2/2)$ for $X\in \{N, R\}$, and deriving $|dz_X/dX|$ from Eq.~\eqref{eq:steady}, we obtain $P(N)$ and $P(R)$ by applying a change of variables: $P(X) = P(z_X) |dz_X/dX|$.
The results yield that $P(N)$ follows a truncated Gaussian distribution, while $P(R)$ follows a more complicated distribution, as shown below:
\begin{widetext}
\begin{equation}
\label{eq:distributions}
\begin{split}
P(N) &= \frac{ h_\text{eff}}{\sqrt{2\pi} \sigma_g} \exp \left[- \frac{(h_\text{eff} N - g_\text{eff})^2}{2 \sigma^2_g } \right]\quad \text{for}~N>0,\\[15pt]
P(R) &= \frac{\gamma^{-1}\epsilon\sigma_R^2 \sigma_C^2 \langle \nu \rangle R^3 + (\sigma_R^2 \mu_K + 2\gamma^{-1}\epsilon \sigma_K^2  \sigma_C^2 \langle \nu \rangle) R + \sigma_K^2 D_\text{eff}}{\sqrt{2\pi}(\sigma^2_K + \sigma^2_R R^2 )^{3/2}} \exp \left[ - \frac{(\gamma^{-1} \epsilon \sigma^2_C \langle \nu \rangle R^2 + D_\text{eff} R - \mu_K)^2}{2(\sigma^2_K + \sigma^2_R R^2 )} \right].
\end{split}
\end{equation}
\end{widetext}
where
\begin{equation}
\label{eq:macro}
\begin{aligned}
h_\text{eff} &= h+\sigma^2_C \langle \chi \rangle,\\
g_\text{eff} &=\epsilon \mu_C\langle R \rangle - \mu_m,\\
\sigma^2_g  &= \sigma^2_m + \epsilon^2\sigma^2_C \langle R^2 \rangle,\\
D_\text{eff} &= \mu_D + \gamma^{-1} \mu_C \langle N \rangle,\\
\sigma^2_R &=\sigma^2_D + \gamma^{-1} \sigma^2_C \langle N^2 \rangle.
\end{aligned}
\end{equation}
For the sake of simplicity, we omit $\epsilon$ in the following equations, as it is fixed to unity.
Detailed derivations are provided in SM~\cite{SM}.

From Eqs.~\eqref{eq:distributions} and \eqref{eq:macro}, we see that the average quantities $\langle N \rangle$, $\langle N^2 \rangle$, $\langle R \rangle$, and $\langle R^2 \rangle$, along with the response functions $\langle\nu\rangle$ and $\langle \chi\rangle$, depend on both $P(N)$ and $P(R)$.
In turn, those distributions $P(N)$ and $P(R)$ themselves depend on these macroscopic quantities.
To resolve this interdependency, we calculate average quantities by solving self-consistency equations.

\begin{figure}[!t]
\centering\includegraphics[width=1.00\linewidth]{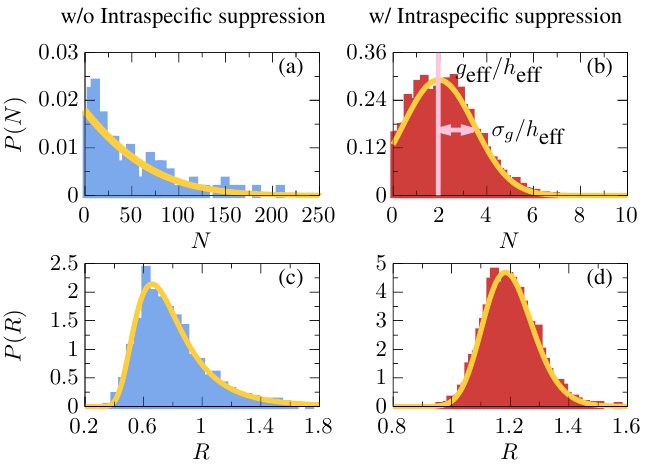}
\caption{
[(a) and (b)] Consumer abundance distribution $P(N)$ for $N > 0$, and [(c) and (d)] resource abundance distribution $P(R)$.
[(a) and (c)] correspond to ecosystems without intraspecific suppression ($h=0$), whereas [(b) and (d)] correspond to those with intraspecific suppression ($h=1/10$).
$P(N)$ follows truncated Gaussian distributions in both cases, with peak position and width determined by the ratios $g_\text{eff}/h_\text{eff}$ and $\sigma_g/h_\text{eff}$, respectively, which is highlighted in (b).
In contrast, $P(R)$ shows skewed distributions in both cases.
Yellow solid lines indicate DMFT results, while colored bars represent numerical results obtained by integrating Eq.~\eqref{eq:model_ext} over $50$ realizations with $S=75$ and $M=50$.
The initial number ratio is fixed at $\gamma = M/S = 2/3$.
The probability of extinct consumers ($N=0$) is not shown.
}
\label{fig:abundances}
\end{figure}

We validate our theoretical results in Eq.~\eqref{eq:distributions} by numerically integrating Eq.~\eqref{eq:model_ext} over $50$ independent realizations with $S=75$ and $M=50$, which gives $\gamma = M/S = 2/3$.
Figure~\ref{fig:abundances} depicts $P(N)$ and $P(R)$ obtained through numerical integration (colored boxes), alongside the distributions derived from DMFT (yellow lines).
Notably, the theoretical and numerical results show excellent agreement.

%
%--------------------
%
%--------------------
\subsection{Relative diversity}
\label{ssec:diversity}
From the consumer abundance distribution $P(N)$, we calculate the relative diversity as
\begin{equation} \label{eq:ErrorFtn}
S^* / M^* = \gamma^{-1} \int_{+0}^\infty P(N) dN = \frac{\gamma^{-1}}{2}\bigg[ 1 + \text{erf} \bigg(\frac{g_\text{eff}}{\sqrt{2} \sigma_g} \bigg) \bigg],
\end{equation}
where $\text{erf}(\cdot)$ denotes the error function.
The relative diversity $S^*/M^*$ exhibits a monotonic increase with $g_\text{eff}/\sigma_g$. 
Thus, the mean growth rate $g_\text{eff}$ and its width $\sigma_g$ together determine $S^*/M^*$.
When $g_\text{eff} = 0$, only half of the consumers survive, as $P(N)$ follows a truncated Gaussian distribution with its peak at $N=g_\text{eff}/h_\text{eff}=0$.
In this case, half of them will have positive abundances by chance.
As $g_\text{eff}$ increases, more consumers attain positive abundances, increasing their chances of survival and leading to higher diversity.
For positive $g_\text{eff}$, however, a larger $\sigma_g$ increases the probability of extinction by producing a fatter tail in $P(N)$ within the range of $N<0$, thereby reducing the diversity.
Conversely, for negative $g_\text{eff}$, the effect of $\sigma_g$ is the opposite; the larger the $\sigma_g$, the more consumers coexist.

\begin{figure}[t]
\centering\includegraphics[width=1.00\linewidth]{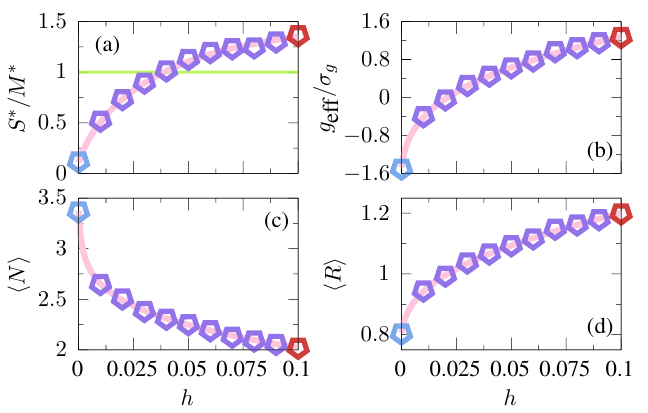}
\caption{
(a) Relative diversity $S^*/M^*$, (b) ratio between the mean to the width of consumers' effective growth rate, $g_\text{eff}/\sigma_g$, (c) mean consumer abundance $\langle N \rangle$, and (d) mean resource abundance $\langle R \rangle$ against intraspecific suppression coefficient $h$.
As $h$ increases, $S^*/M^*$ rises and even exceeds unity, while $\langle N \rangle$ decreases.
In (b), $g_\text{eff}/\sigma_g$ increases with $h$, which explains that the increase in $S^*/M^*$ as derived in Eq.~\eqref{eq:ErrorFtn}.
The pink lines denote the DMFT results, and the symbols represent the numerical results by averaging over $50$ independent realizations with $S=75$ and $M=50$.
Blue and red symbols denote the results for $h=0$ and $h=1/10$, respectively.
Error bars are smaller than the symbol size.
The solid chartreuse line in (a) represents $S^*/M^* = 1$.
Note that the average resource input rate is set to $\mu_K=5$.
}\label{fig:NumberRatio}
\end{figure}

We vary $h$ to investigate the effect of intraspecific suppression on the relative diversity $S^*/M^*$.
As shown in Fig.~\ref{fig:NumberRatio}(a), $S^*/M^*$ increases as $h$ becomes larger.
When intraspecific suppression becomes strong ($h \gtrsim 0.04$), this effect is notably intensified, leading $S^*/M^*$ to exceed $1$.
This indicates that intraspecific suppression can explain the coexistence of more consumer species than the number of available resources.

Such increase of relative diversity $S^*/M^*$ in $h$ is attributed to the rise in $g_\text{eff}/\sigma_g$, as depicted in Fig.~\ref{fig:NumberRatio}(b).
The underlying mechanism is as follows:
Strong intraspecific suppression inhibits the population growth of consumer species, preventing any dominant consumers from emerging, which results in a decrease in the mean consumer abundance $\langle N \rangle$ as $h$ increases [see Fig.~\ref{fig:NumberRatio}(c)].
In consequence, the mean resource abundance $\langle R \rangle$ increases, leading to the increase in $g_\text{eff}$ [see Fig.~\ref{fig:NumberRatio}(d) and Fig.~S7].
Therefore, when intraspecific suppression is strong, numerous consumers can coexist in low abundances, sometimes even exceeding the number of available resources.

%
%--------------------
%--------------------
\subsection{Bounds of the relative diversity}
\label{ssec:bounds}
To assess the extent to which intraspecific suppression expands the range of the relative diversity $S^*/M^*$, we examine the upper and lower bounds.
By differentiating $N$ with respect to $\mu_m$ in Eq.~\eqref{eq:steady}, we derive the relation that $\phi_S = h_\text{eff} \langle\nu\rangle$ (see details in SM~\cite{SM}).
Using the relation along with $S^*/M^* = \gamma^{-1}\phi_S$, we express $S^*/M^*$ in terms of the response functions $\langle\nu\rangle$ and $\langle\chi\rangle$, and $h$ as
\begin{equation}
S^*/M^* = \gamma^{-1} h_\text{eff} \langle\nu\rangle = \gamma^{-1} (h + \sigma^2_c \langle\chi\rangle) \langle\nu\rangle.
\end{equation}
Applying the inequality $0 \leq \gamma^{-1}\sigma_C^2 \langle\chi\rangle \langle\nu\rangle <1/2$ (see details in SM~\cite{SM}), we establish the lower and the upper bounds of $S^*/M^*$ as follows:
\begin{equation}\label{eq:bound_ext}
\gamma^{-1}h \langle\nu\rangle \leq S^*/M^* < \gamma^{-1}h \langle\nu\rangle + 1/2.
\end{equation}
Without intraspecific suppression ($h=0$), $S^*/M^*$ lies between $0$ and $1/2$, as suggested in the previous study~\cite{GCRM_Niche}.
However, both the upper and lower bounds increase by $\gamma^{-1}h \langle\nu\rangle$ with intraspecific suppression ($h>0$), which is always positive.
Consequently, it is possible for the bounds of $S^*/M^*$ to surpass unity with $h$, leading to $S^*/M^*>1$, as observed in Fig.~\ref{fig:NumberRatio}(a).

%--------------------
%--------------------
\subsection{Environmental effect}
\label{ssec:eff_Env}
\begin{figure}[!t]
\centering\includegraphics[width=1.00\linewidth]{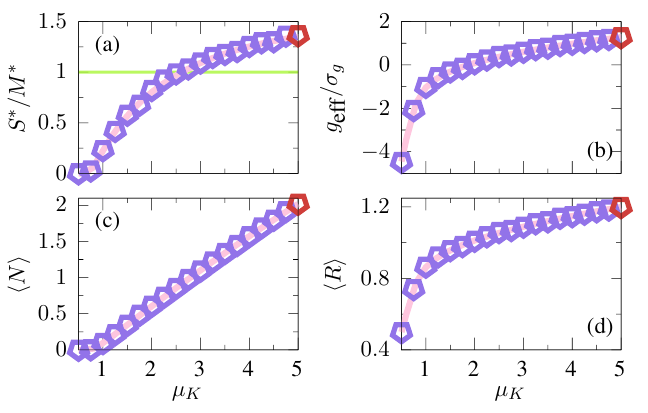}
\caption{
(a) Relative diversity $S^*/M^*$, (b) ratio between the mean to the width of consumers' effective growth rate, $g_\text{eff}/\sigma_g$, (c) mean consumer abundance $\langle N \rangle$, and (d) mean resource abundance $\langle R \rangle$ against average resource input rate $\mu_K$.
In resource-poor environments with low $\mu_K$, $S^*/M^*$ cannot surpass $1$, even with intraspecific suppression.
The initial number ratio to $\gamma=2/3$.
The intraspecific suppression coefficient is set to $h=1/10$.
}\label{fig:eff_Env}
\end{figure}

Environmental conditions play a crucial role in determining whether the number of coexisting species exceeds the number of available resources.
In resource-rich environments (high average resource input rate $\mu_K$), resources are sufficiently abundant, allowing many consumers to access and utilize them for survival.
In comparison, in resource-poor environments (low $\mu_K$), most consumers go extinct, since their minimum energy for survival cannot be satisfied.
Therefore, achieving high relative diversity $S^*/M^*$, even higher than unity, requires resource-rich environments [see Fig.~\ref{fig:eff_Env}(a) and Figs.~S4--S8 in SM~\cite{SM}].
It means that intraspecific suppression has a high impact in resource-rich environments [see left column figures of Figs.~S7--S9 in SM], while the effect is negligible in resource-poor environments [see right column figures of Figs.~S7--S9 in SM].
This is because consumers struggle to grow in resource-poor environments, where individual abundance $N$ is low.
As a result, the effect of intraspecific suppression weakens, as it is proportional to $N^2$.

The increase in $S^*/M^*$ with increasing $\mu_K$ results from the increase of $g_\text{eff}/\sigma_g$ [see Fig.~\ref{fig:eff_Env}(b)].
In contrast to the case where the intraspecific suppression increases, raising $\mu_K$ induces larger mean consumer abundance $\langle N \rangle$ [see Fig.~\ref{fig:eff_Env}(c)].
However, with higher $\mu_K$, resource abundances can remain sufficiently high, even as consumer abundances increase with $\mu_K$, thereby elevating $g_\text{eff}$ [see Fig.~\ref{fig:eff_Env}(d)].

\begin{figure}[t]
\centering\includegraphics[width=\linewidth]{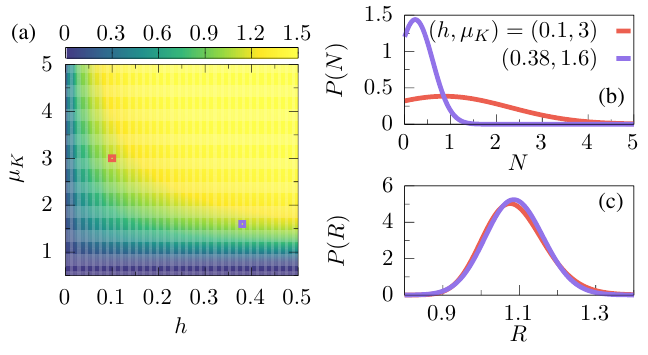}
\caption{
(a) Heatmap of relative diversity $S^*/M^*$ in $(h,\mu_K)$-plane.
(b) $P(N)$ and (c) $P(R)$ at two different points in the plane exhibiting the same $S^*/M^*$.
Vermilion lines in [(b) and (c)] correspond to $(h,\mu_K)=(0.1, 3)$, while purple lines represent $(h,\mu_K)=(0.38, 1.6)$.
Both points exhibit $S^*/M^*\approx 1$, and $g_\text{eff}/\sigma_g \approx 0.6$, yet their distributions $P(N)$ and $P(R)$ differ.
These points are marked with colored box in (a).
$S^*/M^*$ exceeds $1$ only when the environment supplies sufficiently rich resources (large $\mu_K$) with intraspecific suppression [yellow region in (a)].
Without intraspecific suppression ($h=0$), $S^*/M^*$ cannot reach or exceed $1$.
The effect of intraspecific suppression is amplified in resource-rich environments.
}\label{fig:heatmap}
\end{figure}

To take into account both effects of intraspecific suppression and resource input together, we present the heatmap of $S^*/M^*$ in $(h,\mu_K)$-plane in Fig.~\ref{fig:heatmap}(a).
For the relative diversity $S^*/M^*$ to exceed unity, both sufficient suppression and high resource input are required. 
However, while both increased intraspecific suppression and high resource input raise $S^*/M^*$, they have opposite effects on the mean consumer abundances $\langle N \rangle$.
As a result, two parameter sets ($h, \mu_K$) that yield the same $S^*/M^* (\approx 1)$, as marked in Fig.~\ref{fig:heatmap}(a), exhibit different distributions $P(N)$ and $P(R)$ [see Figs.~\ref{fig:heatmap}(b)--(c)].
Since $g_\text{eff}/\sigma_g$ is the key factor determining $S^*/M^*$, the distributions are not necessarily identical for the same $S^*/M^*$.

%
%-----------------------------------
%-----------------------------------
\section{Conclusion and discussion}
\label{sec:conc_disc}
In this paper, we have investigated the effects of intraspecific suppression and resource input on consumer diversity in large ecological systems. 
Employing DMFT, we have derived the consumer abundance distribution $P(N)$, the resource abundance distribution $P(R)$, the formula of the relative diversity $S^*/M^*$, and its upper and lower bounds.
Our theoretical findings have been validated through comparison with numerical integration results.
Notably, our results highlight that the interplay between intraspecific suppression and resource input determines the diversity.
In particular, we find that the number of coexisting consumer species can exceed the number of resource types ($S^/M^ > 1$), challenging a classical principle in ecology and statistical physics.

%%%
Previous studies have explained the emergence of high diversity through metabolic trade-offs, where species face internal constraints in their ability to utilize multiple resources.
For instance, Ref.~\cite{trade_off1} demonstrated that a fixed enzyme budget, modeled as a constraint $\sum_\alpha C_{i\alpha} = E$, can lead to competitive balance and allow the coexistence of more species than resource types.
Similarly, Ref.~\cite{GCRM_Niche} introduced a linear relationship between total resource uptake $\sum_\alpha C_{i\alpha}$ and maintenance cost $m_i$, effectively implementing a trade-off between resource acquisition and survival cost.
In contrast, our model does not include any intrinsic trade-off of this kind.
Both $C_{i\alpha}$ and $m_i$ are sampled independently, and they show no systematic correlation among coexisting species.
Instead, we demonstrate that high diversity can arise from intraspecific suppression, which imposes a nonlinear, abundance-dependent constraint on population growth.
While this mechanism is not derived from an internal resource allocation strategy or evolutionary constraints, it dynamically limits the proliferation of dominant species and promotes stable coexistence.
We interpret this as an emergent ecological constraint that is distinct from classical metabolic trade-offs and can similarly support biodiversity beyond traditional bounds such as the competitive exclusion principle.
%%%

Intraspecific suppression has long been considered in many ecological systems~\cite{SF_MacArthur, SF_Chesson, SF_stability1, SF_stability2}, showing an important role in the stability of the coexistence of consumer and resource species.
However, such studies have focused on two~\cite{SF_stability1} or three~\cite{SF_stability2} species representing each trophic level rather than how many consumer species can survive with respect to the available resources.
Even though it is well known that intraspecific suppression can stabilize the species' coexistence, when combined with environmental factors, quantitatively determining how much intraspecific suppression fosters diversity in large ecological systems with many different consumers and resources has been a nontrivial endeavor.
By integrating intraspecific suppression into GCRM, we have demonstrated that the suppression has the capability to augment biodiversity beyond the constraints of the number of resource kinds, particularly in resource-rich environments.

In our model, interactions between consumers and resources are represented as Gaussian random variables. 
One might therefore expect both the consumer and resource abundance distributions to resemble Gaussian forms.
Indeed, in the case of identical intraspecific suppression ($\sigma_h = 0$), the consumer abundance distribution $P(N)$ follows a truncated Gaussian distribution.
However, the resource abundance distribution $P(R)$ exhibits a markedly different behavior.
When the resource input is sufficiently large, $P(R)$ resembles a lognormal distribution, as shown in Fig.~S2 of SM.
Furthermore, the rank-abundance distribution of $R$ displays a fat tail---a hallmark of lognormal and power-law distributions---highlighted in Fig.~S16 (SM).

The key structural difference between the consumer and resource dynamics lies in the presence of external input to the resources.
This observation suggests that, if consumers were also subject to external influx such as invasion or immigration, their abundance distribution might resemble a lognormal distribution.

Interestingly, even without such external consumer input, introducing heterogeneity in the intraspecific suppression ($\sigma_h \neq 0$) leads to fat-tailed $P(N)$ distributions, as shown in Fig.~S2.
This behavior is not observed under identical suppression, while $P(R)$ remains largely unchanged.
These results underscore the nontrivial effects of internal parameter heterogeneity on abundance patterns, even in a Gaussian interaction framework.
A more comprehensive understanding of how external invasion and internal heterogeneity shape $P(N)$ remains an important direction for future work.

Although $\sigma_h$ modifies the shape of $P(N)$, the relative diversity $S^*/M^*$ shows qualitatively similar behavior between the identical and non-identical intraspecific suppression cases.
We note, however, that the deviation $\sigma_h$ must remain sufficiently small.
Otherwise, the intraspecific suppression term $-h_i N_i^2$ in Eq.~\eqref{eq:model_ext} can become positive for some $h_i < 0$, potentially leading to divergence in $N_i$.

As a natural extension, our analysis of consumer and resource abundance can be adaptable to ecosystems with more complex trophic structures.
For instance, our framework could be extended to systems with three or more levels, including predators, intermediate species that both prey on others and are themselves preyed upon, and externally supplied abiotic resources (see Sec.~V in SM~\cite{SM}).
In this extension, predation may play an important role in biodiversity, serving as a limiting factor~\cite{paradox_resolve}.
Overall, our framework offers a versatile approach for understanding biodiversity across ecologically structured systems, from simple to highly complex food webs.

%-----------------------------------
\section*{Acknowledgments}
The authors acknowledge Deok-Sun Lee, Hyeong-Chai Jeong, and Jong Il Park for fruitful discussion.
This work was supported by a KIAS individual Grant CG091401 at Korea Institute for Advanced Study (S.-G.Y.), an appointment to the YST Program at the APCTP through the Science and Technology Promotion Fund and Lottery Fund of the Korean Government, as well as by the Korean Local Governments --- Gyeongsangbuk-do Province, and Pohang City (S.-G.Y.), and the National Research Foundation of Korea grant funded by the Korea government (MSIT), Grant No. RS-2023-00214071 and RS-2024-00460958 (H.J.P.).
This work was supported by INHA UNIVERSITY Research Grant as well (H.J.P.).
The authors are grateful to the Center for Advanced Computation at KIAS for help with computing resources.
%-----------------------------------
\bibliographystyle{apsrev4-2}

%-----------------------------------
\newpage

%%%%%%%%%%%%%%%%%%%%%%%%%%%%%%%%%%%%%%%%%
%%%%%%%%%%%%%%%%%%%%%%%%%%%%%%%%%%%%%%%%%
\setcounter{equation}{0}
\renewcommand{\thefigure}{S\arabic{figure}}
\renewcommand{\theequation}{S\arabic{equation}}
%%%%%%%%%%%%%%%%%%%%%%%%%%%%%%%%%%%%%%%%%

\begin{CJK*}{UTF8}{}
\title{
Supplemental Material: Interplay between intraspecific suppression and environment in shaping biodiversity
}
%-----------------------------------
\author{Seong-Gyu Yang \CJKfamily{mj}{(양성규)}}
\affiliation{School of Computational Sciences, Korea Institute for Advanced Study, Seoul, 02455, Republic of Korea}
\affiliation{Asia Pacific Center for Theoretical Physics, Pohang, 37673, Republic of Korea}
\author{Hye Jin Park \CJKfamily{mj}{(박혜진)}}
\email{hyejin.park@inha.ac.kr}
\affiliation{Department of Physics, Inha University, Incheon, 22212, Republic of Korea}
%-----------------------------------
%\begin{abstract}

%\end{abstract}
\maketitle
\end{CJK*}

\onecolumngrid
\tableofcontents

\clearpage
\label{supsec:Supple-I}
\section{Table of Variables in Main Text}

\begin{center}
\begin{tabular}{ p{6cm}   p{10cm}  }
\hline
Variable & Description\\
\hline
\hline
$S$  & Initial number of different consumer species  \\
$S^*$  & Number of different consumer species at steady state  \\
$M$  &   Initial number of different resource kinds \\
$M^*$  &   Number of different resource kinds at steady state \\
$\gamma = M/S$  &   Number ratio between initial resource kinds and consumer species \\
\hline
$N_i$  & Abundance of consumer $i$   \\
$R_\alpha$  &   Abundance of resource $\alpha$  \\
\hline
$C_{i\alpha}$  & Consumption rate of resource $\alpha$ by consumer $i$ \\
$h_i$ &  Strength of intraspecific suppression\\
$m_i$ &  Mortality rate of consumer $i$\\
$\epsilon_i$ & Trophic efficiency of consumer $i$\\
$K_\alpha$ &  External input rate of resource $\alpha$\\
$D_\alpha$ & Degradation rate of resource $\alpha$ \\
\hline
$\mu_C$ & Mean value of consumption rate  \\
$\sigma_C$  & Standard deviation of consumption rate \\
$\mu_h ( = h )$ & Mean strength of intraspecific suppression\\
$\sigma_h$ & Standard deviation of intraspecific suppression strength\\
$\mu_m$ &  Mean mortality rate\\
$\sigma_m$ &  Standard deviation of mortality rate\\
$\mu_K$ &  Mean resource input rate\\
$\sigma_K$ &  Standard deviation of resource input rate\\
$\mu_D$ & Mean degradation rate \\
$\sigma_R$ & Standard deviation of degradation rate \\
\hline
$P(N)$ & Consumer abundance distribution \\
$P(R)$ & Resource abundance distribution \\
\hline
$\phi_S = S^*/S = \int_{+0}^{\infty} dN P(N)$ & Surviving probability of consumer species \\
$\langle N \rangle $ &  Mean consumer abundance\\
$\langle N^2 \rangle$ &  Mean squared consumer abundance\\
$\langle R \rangle$ &  Mean resource abundance\\
$\langle R^2 \rangle$ &  Mean squared resource abundance\\
$\langle \nu \rangle= -\langle \partial N / \partial m \rangle$ &  Consumer response function\\
$\langle \chi\rangle = -\langle \partial R / \partial D \rangle$ & Resource response function\\
\hline
$g_\text{eff} = \mu_C\langle R \rangle - \mu_m $ &  Effective consumer growth rate\\
$\sigma_g = \sqrt{\sigma^2_m + \sigma^2_C \langle R^2 \rangle} $ &  Standard deviation of effective growth rate\\
$h_\text{eff} = h + \sigma_C^2 \langle \chi \rangle $ &  Effective intraspecific suppression coefficient\\
$D_\text{eff} = \mu_D + \gamma^{-1}\mu_C \langle N \rangle $ &  Effective resource degradation rate\\
$\sigma_D = \sqrt{\sigma^2_D + \gamma^{-1} \sigma^2_C \langle N^2 \rangle} $ &  Standard deviation of effective degradation rate\\
\hline
\end{tabular}
\end{center}

\clearpage
\label{supsec:Supple-II}
\section{Derivations}
%navigation
\subsection{Cavity method}
\label{supsubsec:Supple-II_A}
In this section, we derive the steady state solutions of consumer and resource abundances using cavity method, following the similar process in Ref.~\cite{Supple-GCRM_Cavity,Supple-GCRM_Niche,Supple-LV_Cavity}.
The consumption rate $C_{i\alpha}$ in 
\begin{equation}
\begin{aligned}
\label{eq:model_ext}
\dot{N}_i &=N_i \left( \epsilon_i\sum_\alpha C_{i\alpha} R_\alpha -m_i -h_i N_i \right),\\
\dot{R}_\alpha &=K_\alpha - \left( D_\alpha +  \sum_i N_i C_{i\alpha} \right)R_\alpha,
\end{aligned}
\end{equation}
has quenched disorder as $C_{i\alpha} = \mu_C/M + z^C_{i\alpha} \sigma_C/\sqrt{M}$, where $z^C_{i\alpha}$ is an independent unit Gaussian random variable.
Other parameters including mortality rate $m_{i} = \mu_m + z^m_i \sigma_m$, intraspecific suppression coefficient $h_{i} = h + z^h_i \sigma_h$, resource input rate $K_{\alpha} = \mu_K + z^K_\alpha \sigma_K$, and degradation rate $D_{\alpha} = \mu_D + z^D_\alpha \sigma_D$ also have quenched disorder, where $z^m_i, z^h_i, z^K_\alpha,$ and $z^D_\alpha$ are unit Gaussian random variables.
For simplicity, the trophic efficiency $\epsilon_i$ of each consumer $i$ is set uniformly to $\epsilon_i=\epsilon=1$ throughout this study.
Separating the parameters into mean and deviation parts, Eq.~\eqref{eq:model_ext} can be rewritten as follows:
\begin{equation}
\begin{aligned}
%\dot{N}_i &= N_i \left( \mu_C\langle R \rangle - \mu_m -h N_i - z^m_i \sigma_m - z^h_i \sigma_h N_i + \frac{\sigma_C}{\sqrt{M}} \sum_\alpha z^C_{i\alpha} R_\alpha  \right),\\
\dot{N}_i &= N_i \left( \epsilon \mu_C\langle R \rangle - \mu_m -h N_i - z^m_i \sigma_m - z^h_i \sigma_h N_i + \epsilon \frac{\sigma_C}{\sqrt{M}} \sum_\alpha z^C_{i\alpha} R_\alpha  \right),\\
\dot{R}_\alpha &= \mu_K + z^K_\alpha \sigma_K - \left( \mu_D +  \gamma^{-1}\mu_C\langle N \rangle + z^D_\alpha \sigma_D + \frac{\sigma_C}{\sqrt{M}}\sum_i z^C_{i\alpha} N_i  \right)R_\alpha,
\end{aligned}
\end{equation}
where $\gamma= M/S$ is the number ratio between consumers and resources, and $\langle N \rangle$ and $\langle R \rangle$ denote the mean consumer abundance and mean resource abundance, respectively.

We consider the immigration of a cavity consumer ($i=0$), and an additional supply of a cavity resource ($\alpha=0$) into a system, assuming that the cavity species and the cavity resource follow the same rule as others.
Those cavities act like small perturbations on the system.
The cavity consumer perturbs the degradation rate as $dD_\alpha = \frac{\sigma_C}{\sqrt{M}}N_0 z^C_{0\alpha}$ and the cavity resource perturbs the consumer's mortality rate as $dm_i = - \epsilon \frac{\sigma_C}{\sqrt{M}}R_0 z^C_{i0}$.
Under those perturbations, we assume that the system linearly responds to those perturbations as
\begin{equation}
\label{eq:Supple-linear_response}
\begin{aligned}
N_i &\approx N_{i \backslash 0} + \sum_{j\backslash 0} \frac{\partial N_i}{\partial m_j} d m_j + \sum_{\beta \backslash 0 } \frac{\partial N_i}{\partial D_\beta} dD_\beta \\
&= N_{i \backslash 0} + \epsilon \frac{\sigma_C}{\sqrt{M}}R_0 \sum_{j\backslash 0} z^C_{j0}\nu^N_{ij} - \frac{\sigma_C}{\sqrt{M}} N_0 \sum_{\beta \backslash 0 } z^C_{0\beta} \chi^N_{i\beta} ,\\[5pt]
R_\alpha &\approx R_{\alpha \backslash 0} + \sum_{k\backslash0} \frac{\partial R_\alpha}{\partial m_k} dm_k + \sum_{\lambda\backslash0}\frac{\partial R_\alpha}{\partial D_\lambda} dD_\lambda \\
&= R_{\alpha \backslash 0} + \epsilon \frac{\sigma_C}{\sqrt{M}}R_0 \sum_{k\backslash0}z^C_{k0}\nu^R_{\alpha k} - \frac{\sigma_C}{\sqrt{M}}N_0 \sum_{\lambda\backslash0} z^C_{0\lambda} \chi^R_{\alpha \lambda},
\end{aligned}
\end{equation}
where $\nu^N_{ij}=-\partial N_i/\partial m_j$, $\nu^R_{\alpha i}=-\partial R_\alpha/\partial m_i$, $\chi^N_{i\alpha}=-\partial N_i/\partial D_\alpha$, and $\chi^R_{\alpha \beta}=-\partial R_\alpha/\partial D_\beta$ are response functions for the perturbations.
To calculate the feedback to the response, we insert Eq.~\eqref{eq:Supple-linear_response} into the equations of motion of cavities.
Since we are dealing with the large ecosystem $(S,  M\gg 1)$, those perturbations from cavities are assumed to not change all macroscopic quantities.
Thus, the equation for cavity consumer and cavity resource can be rewritten as follows:
\begin{equation}
\label{eq:Supple-calculating_cavity}
\begin{aligned}
%\dot{N}_0 &= N_0 \left[ \mu_C\langle R \rangle -\mu_m - h N_0 \vphantom{\frac{\sigma_C}{\sqrt{M}}R_0 \sum_{k\backslash 0} z^C_{k0}\nu^R_{\alpha k}} - z^m_0 \sigma_m - z^h_0 \sigma_h N_0 +\frac{\sigma_C}{\sqrt{M}} z^C_{00} R_0 \right. \\
%& \left.  + \frac{\sigma_C}{\sqrt{M}} \sum_{\alpha \backslash 0} z^C_{0\alpha} \left(R_{\alpha\backslash 0} - \frac{\sigma_C}{\sqrt{M}}N_0 \sum_{\lambda \backslash 0} z^C_{0\lambda }\chi^R_{\alpha \lambda} + \frac{\sigma_C}{\sqrt{M}}R_0 \sum_{k\backslash 0} z^C_{k0}\nu^R_{\alpha k}  \right) \right],
\dot{N}_0 &= N_0 \left[ \epsilon \mu_C\langle R \rangle -\mu_m - h N_0 \vphantom{\frac{\sigma_C}{\sqrt{M}}R_0 \sum_{k\backslash 0} z^C_{k0}\nu^R_{\alpha k}} - z^m_0 \sigma_m - z^h_0 \sigma_h N_0 + \epsilon\frac{\sigma_C}{\sqrt{M}} z^C_{00} R_0 \right. \\
& \left.  + \epsilon \frac{\sigma_C}{\sqrt{M}} \sum_{\alpha \backslash 0} z^C_{0\alpha} \left(R_{\alpha\backslash 0} + \epsilon \frac{\sigma_C}{\sqrt{M}}R_0 \sum_{k\backslash 0} z^C_{k0}\nu^R_{\alpha k}  - \frac{\sigma_C}{\sqrt{M}}N_0 \sum_{\lambda \backslash 0} z^C_{0\lambda }\chi^R_{\alpha \lambda}  \right) \right],
\\[10pt]
\dot{R}_0 &= \mu_K + z^K_0\sigma_K - \left[ \mu_D +  \gamma^{-1}\mu_C\langle N \rangle   \vphantom{\frac{\sigma_C}{\sqrt{M}}N_0\sum_{\beta \backslash 0} z^C_{0\beta} \chi^N_{i\beta}} + z^D_0\sigma_D +\frac{\sigma_C}{\sqrt{M}} z^C_{00} N_0 \right. \\
& \left. + \frac{\sigma_C}{\sqrt{M}} \sum_{i\backslash 0}z^C_{i0} \left( N_{i\backslash 0} + \epsilon \frac{\sigma_C}{\sqrt{M}} R_0 \sum_{j\backslash 0} z^C_{j0} \nu^N_{ij} - \frac{\sigma_C}{\sqrt{M}}N_0\sum_{\beta \backslash 0} z^C_{0\beta} \chi^N_{i\beta} \right) \right] R_0.
\end{aligned}
\end{equation}
Utilizing the property of independent random variables, the last terms in Eq.~\eqref{eq:Supple-calculating_cavity} go to zero.
Ignoring those terms, we get the equations that cavity consumer and resource should follow at a steady state,
\begin{equation}
\label{eq:Supple-cavity_results}
\begin{aligned}
%0 &= N \left[ \mu_C\langle R \rangle - \mu_m - (h +\sigma^2_C \chi) N + z_N \sqrt{\sigma^2_m + \sigma^2_h N^2 + \sigma^2_C \langle R^2 \rangle} \right],
%\\
%0 &= -\gamma^{-1}\sigma^2_C\nu R^2 - \left( \mu_D +  \gamma^{-1}\mu_C\langle N \rangle \right)R + \mu_K + z_R \sqrt{\sigma^2_K + (\sigma^2_D + \gamma^{-1}\sigma^2_C \langle N^2 \rangle ) R^2},
0 &= N \left[ \epsilon \mu_C\langle R \rangle - \mu_m - (h +\sigma^2_C \langle \chi \rangle) N + z_N \sqrt{\sigma^2_m + \sigma^2_h N^2 + \epsilon^2 \sigma^2_C \langle R^2 \rangle} \right],
\\
0 &= -\gamma^{-1}\epsilon\sigma^2_C \langle \nu \rangle R^2 - \left( \mu_D +  \gamma^{-1}\mu_C\langle N \rangle \right)R + \mu_K + z_R \sqrt{\sigma^2_K + (\sigma^2_D + \gamma^{-1}\sigma^2_C \langle N^2 \rangle ) R^2},
\end{aligned}
\end{equation}
where $\langle \nu \rangle= \frac{1}{S}\sum_i \nu^N_{ii}$ and $\langle\chi\rangle = \frac{1}{M} \sum_\alpha \chi^R_{\alpha\alpha}$, and $z_N$ and $z_R$ are independent unit Gaussian random variables.
Since all species behave in the same way as the cavity species, we omit the subscript $0$ in Eq.~\eqref{eq:Supple-cavity_results}.

%%%%%%%%%%%%%%%%%%%%%%%%%%%%%%%%%%%%%%%%%%%%%%%%%%%
\subsection{Generating functional analysis}
\label{supsubsec:Supple-II_B}
Although the cavity method is a intuitive and useful, it has somehow lack of mathematical rigor, especially in taking average over ensemble and disorder.
In this reason, we validate the result of Eq.~\eqref{eq:Supple-cavity_results} using generating functional analysis, following the similar procedure described in Ref.~\cite{Supple-LV_DMFT,Supple-GCRM_DMFT}.
The generating functional can be written as follows:
\begin{equation}
\label{eq:Supple-generating_functional}
Z(\bm{\psi}, \bm{\phi}) =\overline{
\left\langle \exp \left( i \sum_j \int dt\, \psi_j(t) N_j(t) + i \sum_\alpha \int dt\, \phi_\alpha(t)R_\alpha(t)\right) \right\rangle},
\end{equation}
where $\bm{\psi}$, and $\bm{\phi}$ are auxiliary fields.
Here, $\langle \cdots \rangle $ and $\overline{\cdots}$ denote the path and the disorder averages, respectively.
Inside Eq.~\eqref{eq:Supple-generating_functional}, $N_j(t)$ and $R_\alpha(t)$ should follow Eq.~\eqref{eq:model_ext}.
We apply the time-dependent response probing fields in Eq.~\eqref{eq:model_ext}, and the path average is written as
\begin{equation}
\label{eq:Supple-generating_functional2}
\begin{aligned}
%Z(\bm{\psi}, \bm{\phi})  &=\overline{
%\int \mathscr{D}[\bm{N}] \mathscr{D}[\bm{R}] \exp\left( i\sum_j\int dt\, \psi_j N_j \right) \exp\left( i \sum_\alpha \int dt\, \phi_\alpha R_\alpha \right) }\\
%& \overline{\times \prod_{t,j}\delta \left[ \frac{\dot{N}_j}{N_j} - \sum_\alpha C_{j\alpha} R_\alpha + m_j + \eta_j^m + \left(h_j + \eta_j^h\right)N_j \right] } \\
%&\quad \overline{
%\times  \prod_{t,\alpha}\delta \left( \frac{\dot{R}_\alpha - K_\alpha - \eta_\alpha^K}{R_\alpha}  + D_\alpha +\eta_\alpha^D + \sum_j C_{j\alpha} N_j \right) },
Z(\bm{\psi}, \bm{\phi})  &=\overline{
\int \mathscr{D}[\bm{N}] \mathscr{D}[\bm{R}] \exp\left( i\sum_j\int dt\, \psi_j N_j \right) \exp\left( i \sum_\alpha \int dt\, \phi_\alpha R_\alpha \right) }\\
& \overline{\times \prod_{t,j}\delta \left[ \frac{\dot{N}_j}{N_j} - \epsilon \sum_\alpha C_{j\alpha} R_\alpha + m_j + \eta_j^m + \left(h_j + \eta_j^h\right)N_j \right] } \\
&\quad \overline{
\times  \prod_{t,\alpha}\delta \left( \frac{\dot{R}_\alpha - K_\alpha - \eta_\alpha^K}{R_\alpha}  + D_\alpha +\eta_\alpha^D + \sum_j C_{j\alpha} N_j \right) },
\end{aligned}
\end{equation}
where $\bm{\eta}^m(t)$, $\bm{\eta}^h(t)$, $\bm{\eta}^D(t)$, and $\bm{\eta}^K(t)$ are the response probing fields.
Utilizing the property of the delta function and separating it into mean and disorder parts, we rewrite Eq.~\eqref{eq:Supple-generating_functional2} as
\begin{equation}
\begin{aligned}
%Z(\bm{\psi}, \bm{\phi}) &= \int \mathscr{D}[\bm{N},\bm{\hat{N}}] \mathscr{D}[\bm{R},\bm{\hat{R}}] \exp \left(i\sum_j\int dt\, \psi_j N_j \right) \exp \left(i\sum_\alpha \int dt\, \phi_\alpha R_\alpha \right) \\
%&\times \exp\left[i \sum_j \int dt\, \hat{N}_j \left( \frac{\dot{N}_j}{N_j} - \frac{\mu_C}{M}\sum_\alpha R_\alpha + m_j + \eta_j^m + (h_j + \eta_j^h)N_j \right) \right] \\
%&\times \exp \left[i \sum_\alpha \int dt\, \hat{R}_\alpha \left( \frac{\dot{R}_\alpha-K_\alpha-\eta_\alpha^K}{R_\alpha}  +D_\alpha +\eta_\alpha^D + \gamma\frac{\mu_C}{S}\sum_j N_j \right) \right] \\
%&\times \overline{ \exp \left[i \sum_j \int dt\, \hat{N}_j\left( - \frac{\sigma_C}{\sqrt{M}} \sum_\alpha R_\alpha z_{i\alpha}^C  + \sigma_m z_j^m + \sigma_h z_j^h N_j \right) \right] } \\
%&\overline{ \times \exp \left[i \sum_\alpha \int dt\, \hat{R}_\alpha \left( \frac{\sigma_C}{\sqrt{M}} \sum_j  N_j z_{j\alpha}^C + \sigma_D z_\alpha^D - \frac{\sigma_K}{R_\alpha}z_\alpha^K  \right)\right]},
Z(\bm{\psi}, \bm{\phi}) &= \int \mathscr{D}[\bm{N},\bm{\hat{N}}] \mathscr{D}[\bm{R},\bm{\hat{R}}] \exp \left(i\sum_j\int dt\, \psi_j N_j \right) \exp \left(i\sum_\alpha \int dt\, \phi_\alpha R_\alpha \right) \\
&\times \exp\left[i \sum_j \int dt\, \hat{N}_j \left( \frac{\dot{N}_j}{N_j} - \epsilon \frac{\mu_C}{M}\sum_\alpha R_\alpha + m_j + \eta_j^m + (h_j + \eta_j^h)N_j \right) \right] \\
&\times \exp \left[i \sum_\alpha \int dt\, \hat{R}_\alpha \left( \frac{\dot{R}_\alpha-K_\alpha-\eta_\alpha^K}{R_\alpha}  +D_\alpha +\eta_\alpha^D + \gamma\frac{\mu_C}{S}\sum_j N_j \right) \right] \\
&\times \overline{ \exp \left[i \sum_j \int dt\, \hat{N}_j\left( - \epsilon \frac{\sigma_C}{\sqrt{M}} \sum_\alpha R_\alpha z_{i\alpha}^C  + \sigma_m z_j^m + \sigma_h z_j^h N_j \right) \right] } \\
&\overline{ \times \exp \left[i \sum_\alpha \int dt\, \hat{R}_\alpha \left( \frac{\sigma_C}{\sqrt{M}} \sum_j  N_j z_{j\alpha}^C + \sigma_D z_\alpha^D - \frac{\sigma_K}{R_\alpha}z_\alpha^K  \right)\right]},
\end{aligned}
\end{equation}
where $z^C$, $z^m$, $z^h$, $z^D$, and $z^K$ are unit Gaussian random variables.
Then, we do not need to take the disorder average for all terms but only for the terms with disorders.
After averaging over disorder utilizing the cumulant expansion for the Gaussian random variables, we obtain the generating functional as
\begin{equation}
\label{eq:Supple-generating_functional3}
\begin{aligned}
%Z(\bm{\psi}, \bm{\phi}) & = \int \mathscr{D}[\bm{N},\bm{\hat{N}}] \mathscr{D}[\bm{R},\bm{\hat{R}}]\exp \left(i\sum_j\int dt\, \psi_j N_j \right) \exp \left(i\sum_\alpha \int dt\, \phi_\alpha R_\alpha \right) \\
%&\times \exp\left[i \sum_j \int dt\, \hat{N}_j\left( \frac{\dot{N}_j}{N_j} - \frac{\mu_C}{M}\sum_\alpha R_\alpha + m_j + \eta_j^m + \left(h_j + \eta_j^h \right)N_j \right) \right] \\
%&\times \exp \left[i \sum_\alpha \int dt\, \hat{R}_\alpha \left( \frac{\dot{R}_\alpha-K_\alpha-\eta_\alpha^K}{R_\alpha}  +D_\alpha +\eta_\alpha^D + \gamma\frac{\mu_C}{S}\sum_j N_j \right) \right] \\
%&\times \exp \left[ -\frac{\sigma_m^2}{2}\sum_j \iint dt dt'\, \hat{N}_j\hat{N}'_j \right] \exp \left[ -\frac{\sigma_D^2}{2}\sum_\alpha \iint dt dt'\, \hat{R}_\alpha \hat{R}'_\alpha \right]\\
%&\times \exp \left[- \frac{\sigma_h^2}{2}\sum_j \iint dtdt'\, N_j\hat{N}_j N'_j\hat{N}'_j \right]\\
%&\times \exp \left[ -\frac{\sigma_K^2}{2}\sum_\alpha \iint dtdt'\, \frac{\hat{R}_\alpha %\hat{R}'_\alpha}{R_\alpha R'_\alpha} \right]\\
%&\times \exp \left[-\frac{\sigma_C^2}{2M} \sum_{j,\alpha}\iint dtdt'\, \left( N_j N'_j \hat{R}_\alpha \hat{R}'_\alpha + \hat{N}_j\hat{N}'_j R_\alpha R'_\alpha - 2 N_j \hat{N}'_j R'_\alpha \hat{R}_\alpha  \right) \right],
Z(\bm{\psi}, \bm{\phi}) & = \int \mathscr{D}[\bm{N},\bm{\hat{N}}] \mathscr{D}[\bm{R},\bm{\hat{R}}]\exp \left(i\sum_j\int dt\, \psi_j N_j \right) \exp \left(i\sum_\alpha \int dt\, \phi_\alpha R_\alpha \right) \\
&\times \exp\left[i \sum_j \int dt\, \hat{N}_j\left( \frac{\dot{N}_j}{N_j} - \epsilon \frac{\mu_C}{M}\sum_\alpha R_\alpha + m_j + \eta_j^m + \left(h_j + \eta_j^h \right)N_j \right) \right] \\
&\times \exp \left[i \sum_\alpha \int dt\, \hat{R}_\alpha \left( \frac{\dot{R}_\alpha-K_\alpha-\eta_\alpha^K}{R_\alpha}  +D_\alpha +\eta_\alpha^D + \gamma\frac{\mu_C}{S}\sum_j N_j \right) \right] \\
&\times \exp \left[ -\frac{\sigma_m^2}{2}\sum_j \iint dt dt'\, \hat{N}_j\hat{N}'_j \right] \exp \left[ -\frac{\sigma_D^2}{2}\sum_\alpha \iint dt dt'\, \hat{R}_\alpha \hat{R}'_\alpha \right]\\
&\times \exp \left[- \frac{\sigma_h^2}{2}\sum_j \iint dtdt'\, N_j\hat{N}_j N'_j\hat{N}'_j \right] \exp \left[ -\frac{\sigma_K^2}{2}\sum_\alpha \iint dtdt'\, \frac{\hat{R}_\alpha \hat{R}'_\alpha}{R_\alpha R'_\alpha} \right]\\
&\times \exp \left[-\frac{\sigma_C^2}{2M} \sum_{j,\alpha}\iint dtdt'\, \left( N_j N'_j \hat{R}_\alpha \hat{R}'_\alpha + \epsilon^2 \hat{N}_j\hat{N}'_j R_\alpha R'_\alpha - 2 \epsilon N_j \hat{N}'_j R'_\alpha \hat{R}_\alpha  \right) \right],
\end{aligned}
\end{equation}
where the variables $X= X(t)$, and $X'= X(t')$.
From Eq.~\eqref{eq:Supple-generating_functional3}, we obtain macroscopic quantities by derivative with field variables
\begin{equation}
\begin{aligned}
\rho_N(t) &= \frac{1}{S}\sum_j N_j(t) = -i \frac{1}{S} \sum_j \left. \frac{\partial Z(\bm{\varphi})}{\partial \psi_j(t)} \right|_{\bm{\varphi}= 0}, \\
\rho_R(t) &= \frac{1}{M}\sum_\alpha R_\alpha(t) = -i \frac{1}{M} \sum_\alpha \left. \frac{\partial Z(\bm{\varphi})}{\partial \phi_\alpha (t)} \right|_{\bm{\varphi}= 0},\\
\lambda_N(t) &= \frac{1}{S}\sum_j \hat{N}_j(t) = -i \frac{1}{S}\sum_j \left. \frac{\partial Z(\bm{\varphi})}{\partial \eta^m_j(t)} \right|_{\bm{\varphi}= 0}, \\
\lambda_R(t) &= \frac{1}{M}\sum_\alpha \hat{R}_\alpha(t) = -i \frac{1}{M} \sum_\alpha \left. \frac{\partial Z(\bm{\varphi})}{\partial \eta^D_\alpha(t)} \right|_{\bm{\varphi}= 0}, \\
C_N(t,t') &= \frac{1}{S}\sum_j N_j(t) N_j(t') = - \frac{1}{S}\sum_j \left. \frac{\partial^2 Z(\bm{\varphi})}{\partial \psi_j(t) \partial \psi_j(t')} \right|_{\bm{\varphi}= 0}, \\
C_R(t,t') &= \frac{1}{M}\sum_\alpha R_\alpha(t) R_\alpha(t') = -\frac{1}{M}\sum_\alpha \left. \frac{\partial^2 Z(\bm{\varphi})}{\partial \phi_\alpha(t) \partial \phi_\alpha(t')} \right|_{\bm{\varphi}= 0}, \\
L_N(t,t') &= \frac{1}{S}\sum_j \hat{N}_j(t)\hat{N}_j(t') = -\frac{1}{S}\sum_j \left. \frac{\partial^2 Z(\bm{\varphi})}{\partial \eta^m_j(t) \partial \eta^m_j(t')} \right|_{\bm{\varphi}= 0}, \\
L_R(t,t') &= \frac{1}{M}\sum_\alpha \hat{R}_\alpha(t) \hat{R}_\alpha(t') = -\frac{1}{M}\sum_\alpha \left. \frac{\partial^2 Z(\bm{\varphi})}{\partial \eta^D_\alpha(t) \partial \eta^D_\alpha(t')} \right|_{\bm{\varphi}= 0}, \\
K_N(t,t') &= \frac{1}{S}\sum_j N_j(t)\hat{N}_j(t') = -\frac{1}{S}\sum_j \left. \frac{\partial^2 Z(\bm{\varphi})}{\partial \psi_j(t) \partial \eta^m_j(t')} \right|_{\bm{\varphi}= 0}, \\
K_R(t,t') &= \frac{1}{M}\sum_\alpha R_\alpha(t)\hat{R}_\alpha(t') = -\frac{1}{M}\sum_\alpha \left. \frac{\partial^2 Z(\bm{\varphi})}{\partial \phi_\alpha(t) \partial \eta^D_\alpha(t')} \right|_{\bm{\varphi}= 0}, \\
U_N(t,t') &= \frac{1}{S}\sum_j N_j(t)N_j(t')\hat{N}_j(t)\hat{N}_j(t') = -\frac{1}{S}\sum_j \left. \frac{\partial^2 Z(\bm{\varphi})}{\partial \eta^h_h(t) \partial \eta^h_j(t')} \right|_{\bm{\varphi}= 0}, \\
V_R(t,t') &= \frac{1}{M}\sum_\alpha \frac{\hat{R}_\alpha(t)\hat{R}_\alpha(t')}{R_\alpha(t)R_\alpha(t')} = - \frac{1}{M}\sum_\alpha \left. \frac{\partial^2 Z(\bm{\varphi})}{\partial \eta^K_\alpha(t) \partial \eta^K_\alpha(t')} \right|_{\bm{\varphi}= 0},
\end{aligned}
\end{equation}
where $\bm{\varphi}= (\bm{\psi}, \bm{\phi})$.
We can easily find that $\lambda_N= \lambda_R = L_N = L_R = U_N = V_R = 0$, because $Z(\bm{\varphi} =0 ) = 1$.

We define 
$\bm{\Pi} = (\rho_N,\rho_R,\lambda_N,\lambda_R,C_N,C_R,L_N,L_R,K_N,K_R,U_N,V_R)$ to write the generating functional more shortly, and the generating functional is rewritten in terms of macroscopic quantities as follows:
\begin{equation}
\label{eq:Supple-Z}
Z(\bm{\varphi}) =  \int\mathscr{D}[\bm{\Pi},\hat{\bm{\Pi}}]
\exp\left[ S \left(\Phi(\bm{\Pi}) +\Psi(\bm{\Pi}, \bm{\hat{\Pi}}) +  \Omega_N(\hat{\bm{\Pi}}) + \Omega_R(\hat{\bm{\Pi}}) \right) \right],
\end{equation}
where
\begin{equation}
\begin{aligned}
&\quad\Phi(\bm{\Pi}) = i \mu_C \int dt\, (\lambda_R\rho_N - \epsilon \lambda_N\rho_R) \\
&\qquad -\frac{\sigma_C^2}{2}\iint dt dt'\, \left[C_N(t,t')L_R(t,t') + \epsilon^2 L_N(t,t')C_R(t,t') - 2 \epsilon K_N(t,t')K_R(t',t) \right] \\
&\qquad -\frac{1}{2}\iint dt dt'\, \left[ \sigma_m^2 L_N(t,t') 
+ \gamma \sigma_D^2 L_R(t,t')  
+ \sigma_h^2 U_N(t,t') + \gamma \sigma_K^2 V_R(t,t') \right],
\end{aligned}
\end{equation}

\begin{equation}
\begin{aligned}
\Psi(\bm{\Pi}, \bm{\hat{\Pi}}) &= i\int dt\, \left(\hat{\rho}_N\rho_N + \hat{\lambda}_N\lambda_N + \gamma \hat{\rho}_R\rho_R + \gamma \hat{\lambda}_R\lambda_R \right)\\
&\quad + i \iint dt dt' \, \left[\hat{C}_N C_N + \hat{L}_N L_N + \hat{K}_N K_N + \hat{U}_N U_N  \right]\\
&\quad +  i \gamma \iint dt dt' \, \left[ \hat{C}_R C_R + \hat{L}_R L_R +  \hat{K}_R K_R + \hat{V}_R V_R \right],
\end{aligned}
\end{equation}

\begin{equation}
\begin{aligned}
\Omega_N(\bm{\hat{\Pi}}) & = \frac{1}{S} \sum_j \ln \left\{ \int \mathscr{D}[N_j,\hat{N}_j] \exp \left(i\int dt\, \psi_j N_j \right) \right. \\
&\quad \times \left. \exp \left[ i  \int dt\, \hat{N}_j\left( \frac{\dot{N}_j}{N_j} + m_j + h_j N_j + \eta_j^m + \eta_j^h N_j \right)  \right.\right. \\
&\quad -i \int dt \left(\hat{\rho}_N  N_j + \hat{\lambda}_N \hat{N}_j \right) - i\iint dt dt'\, \hat{U}_N N_j N'_j\hat{N}_j\hat{N}'_j   \\
&\quad  \left. \left. \vphantom{\int dt\, \hat{N}_j \frac{\dot{N}_j}{N_j}} -i\iint dt dt'\, \left(\hat{C}_N N_j N'_j + \hat{L}_N \hat{N}_j\hat{N}'_j + \hat{K}_N N_j\hat{N}'_j \right) \right] \right\},
\end{aligned}
\end{equation}
and
\begin{equation}
\label{eq:Supple-Fields}
\begin{aligned}
\Omega_R(\bm{\hat{\Pi}}) &= \frac{1}{S} \sum_\alpha \ln\left\{ \int \mathscr{D}[R_\alpha, \hat{R}_\alpha] \exp \left( i \int dt\, \phi_\alpha R_\alpha \right) \right.\\
& \quad \times \left. \exp \left[i  \int dt\, \hat{R}_\alpha \left( \frac{\dot{R}_\alpha-K_\alpha-\eta_\alpha^K}{R_\alpha} + D_\alpha +\eta_\alpha^D \right) \right.\right. \\
&\quad -i\int dt\, \left( \hat{\rho}_R R_\alpha + \hat{\lambda}_R \hat{R}_\alpha \right) -i\iint dt dt'\, \hat{V}_R \frac{\hat{R}_\alpha \hat{R}'_\alpha}{R_\alpha R'_\alpha}  \\
&\quad \left.\left. \vphantom{\int dt\, \hat{N}_j \frac{\dot{N}_j}{N_j}} -i \iint dtdt'\, \left( \hat{C}_R R_\alpha R'_\alpha + \hat{L}_R \hat{R}_\alpha \hat{R}'_\alpha + \hat{K}_R  R_\alpha \hat{R}'_\alpha \right) \right] \right\}.
\end{aligned}
\end{equation}

Using the saddle-point approximation, we can evaluate Eq.~\eqref{eq:Supple-Z} as $Z[\bm{\varphi}] \approx \exp[S(\Phi^* + \Psi^* + \Omega_N^* + \Omega_R^*)]$, where $\Phi^*$, $\Psi^*$, $\Omega_N^*$, and $\Omega_R^*$ satisfy $\partial_{\bm{\Pi}} (\Phi^* + \Psi^*) = 0$ and $\partial_{\bm{\hat{\Pi}}} (\Psi^* + \Omega_N^* + \Omega_R^*) = 0$.
From the first condition, we get the relations of
\begin{equation}
\label{eq:Supple-relations_GFA}
\begin{aligned}
\hat{\rho}_N &= \hat{\rho}_R = \hat{C}_N = \hat{C}_R = 0, \\
\hat{\lambda}_N &= \mu_C\rho_R, \\
\gamma \hat{\lambda}_R &= -\mu_C\rho_N, \\
i\hat{L}_N &= \epsilon^2\frac{\sigma_C^2}{2} C_R + \frac{\sigma_m^2}{2}, \\
i\gamma \hat{L}_R &= \frac{\sigma_C^2}{2} C_N + \gamma \frac{\sigma_D^2}{2}, \\
i\hat{K}_N &= - \epsilon \sigma_C^2 K_R(t',t), \\
i\gamma \hat{K}_R &= - \epsilon \sigma_C^2 K_N(t',t), \\
i\hat{U}_N &= \frac{\sigma_h^2}{2}, \\
i\hat{V}_R &= \frac{\sigma_K^2}{2}.
\end{aligned}
\end{equation}
In addition, from the second condition, the macroscopic quantities are rewritten as
\begin{equation}
\label{eq:Supple-averages_GFA}
\begin{aligned}
\rho_N(t) &= \left\langle \frac{1}{S} \sum_i N_i (t)\right\rangle_\star, \\
\rho_R(t) &= \left\langle \frac{1}{M} \sum_\alpha R_\alpha (t)\right\rangle_\star, \\
\lambda_N(t) &= \left\langle \frac{1}{S} \sum_i \hat{N}_i (t)\right\rangle_\star, \\
\lambda_R(t) &= \left\langle \frac{1}{M} \sum_\alpha \hat{R}_\alpha (t) \right\rangle_\star, \\
C_N(t,t') &= \left\langle \frac{1}{S} \sum_i N_i(t) N_i(t') \right\rangle_\star, \\
C_R(t,t') &= \left\langle \frac{1}{M} \sum_\alpha R_\alpha(t) R_\alpha(t') \right\rangle_\star, \\
L_N(t,t') &= \left\langle \frac{1}{S} \sum_i \hat{N}_i(t) \hat{N}_i(t') \right\rangle_\star, \\
L_R(t,t') &= \left\langle \frac{1}{M} \sum_\alpha \hat{R}_\alpha(t) \hat{R}_\alpha(t') \right\rangle_\star, \\
K_N(t,t') &= \left\langle \frac{1}{S} \sum_i N_i(t) \hat{N}_i(t') \right\rangle_\star, \\
K_R(t,t') &= \left\langle \frac{1}{M} \sum_\alpha R_\alpha(t) \hat{R}_\alpha(t') \right\rangle_\star, \\
U_N(t,t') &= \left\langle \frac{1}{S} \sum_i N_i(t) N_i(t') \hat{N}_i(t) \hat{N}'_i(t') \right\rangle_\star, \\
V_R(t,t') &= \left\langle \frac{1}{M} \sum_\alpha \frac{\hat{R}_\alpha(t) \hat{R}_\alpha(t')}{R_\alpha(t) R_\alpha(t')}  \right\rangle_\star,
\end{aligned}
\end{equation}
where $\langle A \rangle_\star = \int \mathscr{D}[\bm{N},\hat{\bm{N}}] \mathscr{D}[\bm{R},\hat{\bm{R}}]\, \left. A \exp(\cdots)\, \right/ \int \mathscr{D}[\bm{N},\hat{\bm{N}}] \mathscr{D}[\bm{R},\hat{\bm{R}}] \exp(\cdots)$.
Using the relations in Eq.~\eqref{eq:Supple-relations_GFA} and the expressions of macroscopic quantities in Eq.~\eqref{eq:Supple-averages_GFA}, $\Omega^*_N$ and $\Omega^*_R$ are rewritten as follows:
\begin{equation}
\label{eq:Supple-Omega_N}
\begin{aligned}
\Omega^*_N & = \ln \left\{ \int \mathscr{D}[N,\hat{N}] \exp \left(i\int dt\, \psi(t) N(t) \right) \right. \\
&\quad \times \exp \left[i \int dt\, \hat{N}(t)\left( \frac{\dot{N}(t)}{N(t)} - \epsilon \mu_C \rho_R(t) + \mu_m + h N(t) \right.\right. \\
& \qquad\quad \left. \left. - i\epsilon \sigma_C^2 \int dt'\, K_R(t,t')N(t') + \eta^m(t) + \eta^h(t) N(t) \right) \right] \\
&\quad \left. \times \exp \left[- \frac{1}{2} \iint dtdt'\, \hat{N}(t)\hat{N}(t') \left( \textcolor{myred}{\epsilon^2 \sigma_C^2C_R(t,t') + \sigma_m^2 + \sigma_h^2 N(t) N(t')} \right) \right] \right\} ,
\end{aligned}
\end{equation}

\begin{equation}
\label{eq:Supple-Omega_R}
\begin{aligned}
\Omega^*_R &= \gamma \ln\left\{ \int \mathscr{D}[R,\hat{R}] \exp \left( i \int dt\, \phi(t) R(t) \right) \right.\\
&\quad \times \exp \left[i \int dt\, \hat{R}(t) \left( \frac{\dot{R}(t) - \mu_K -\eta^K(t)}{R(t)} + \mu_D + \gamma^{-1}\mu_C \rho_N(t) \right.\right.\\
&\qquad\quad \left.\left.- i \gamma^{-1} \epsilon \sigma_C^2 \int dt'\, K_N(t,t')R(t') + \eta^D(t) \right) \right]\\
&\quad \left. \times \exp \left[- \frac{1}{2} \iint dtdt'\, \hat{R}(t) \hat{R}(t') \left(\textcolor{myblue}{ \gamma^{-1}\sigma_C^2C_N(t,t') + \sigma_D^2 + \frac{\sigma_K^2}{R(t)R(t')}} \right)\right] \right\}.
\end{aligned}
\end{equation}

Now, we have the effective mean-field dynamics of consumer abundance from Eq.~\eqref{eq:Supple-Omega_N} and resource abundance from Eq.~\eqref{eq:Supple-Omega_R} as
\begin{equation}
\label{eq:Supple-effective_dyn}
\begin{aligned}
\dot{N}(t) &= N(t) \left( \epsilon \mu_C\rho_R(t) - m(t) - h(t) N(t) \vphantom{i \sigma_C^2 \int dt'\, K_R(t,t')N(t')} \right.\\
&\qquad\quad \left. + i \epsilon^2 \sigma_C^2 \int dt'\, K_R(t,t')N(t') + \xi^N(t) + \zeta^N(t) N(t) \right), \\[10pt]
\dot{R}(t) &= K(t) + \zeta^R(t) - \left( D(t) +\gamma^{-1}\mu_C \rho_N(t) \vphantom{i\gamma^{-1}\sigma_C^2 \int dt'\, K_N(t,t')R(t')}\right.\\
&\qquad\quad \left.- i\gamma^{-1} \epsilon \sigma_C^2 \int dt'\, K_N(t,t')R(t') + \xi^R(t) \right) R(t),
\end{aligned}
\end{equation}
where $m(t) = \mu_m + \eta^m(t)$, $h(t) = h + \eta^h(t)$, $K(t) = \mu_K + \eta^K(t)$, $D(t) = \mu_D + \eta^D(t)$.
Here, $\xi^N$ and $\xi^R$ are noises with $\langle \xi^N(t) \xi^N(t') \rangle = \epsilon^2 \sigma_C^2C_R(t,t') + \sigma_m^2$ and $\langle \xi^R(t)\xi^R(t') \rangle = \gamma^{-1}\sigma_C^2 C_N(t,t') + \sigma_D^2$, and $\zeta^N$ and $\zeta^R$ are noises with $\langle \zeta^N(t) \zeta^N(t') \rangle = \sigma_h^2 $ and $\langle \zeta^R(t) \zeta^R(t') \rangle = \sigma_K^2 $.
We rewrite the effective mean-field dynamics of Eq.~\eqref{eq:Supple-effective_dyn} for zero response probing fields with defining $G_R(t,t') = - i K_R(t,t') = - \left\langle \left. \frac{\partial R(t)}{\partial D(t')} \right |_{D(t')=D} \right\rangle_\star$ and $G_N(t,t') = - i K_N(t,t') = - \left\langle \left. \frac{\partial N(t)}{\partial m(t')}  \right |_{m(t') =m}\right\rangle_\star$ as
\begin{equation}
\label{eq:Supple-effective_dyn2}
\begin{aligned}
\dot{N}(t) &= N(t) \left( \epsilon \mu_C\rho_R(t) - \mu_m - h N(t) + \xi^N(t) + \zeta^N(t) N(t) - \epsilon^2 \sigma_C^2 \int dt'\, G_R(t,t')N(t') \right),\\[5pt]
\dot{R}(t) &= \mu_K  + \zeta^R(t) - \left( \mu_D +\gamma^{-1}\mu_C \rho_N(t) + \xi^R(t)+ \gamma^{-1}\epsilon \sigma_C^2 \int dt'\, G_N(t,t')R(t') \right) R(t).
\end{aligned}
\end{equation}

At the steady state $(t \rightarrow \infty)$, we assume that there are no long-term memories, then we can rewrite the response functions as functions of time difference as $G_X(t,t')=G_X(\tau)$ $(X=N \text{ or } R)$ where $\tau=t-t'$.
In this state, $\rho_N=\langle N \rangle$ and $\rho_R = \langle R\rangle$, and the correlation functions become $C_N = \langle N^2 \rangle$ and $C_R= \langle R^2 \rangle$.
We define the response functions as $\langle \nu \rangle = \int d\tau\, G_N(\tau)$ and $\langle \chi \rangle= \int d\tau\, G_R(\tau)$, and rewrite Eq.~\eqref{eq:Supple-effective_dyn2} as follows:
\begin{equation}
\begin{aligned}
0 &= N \left[ \epsilon \mu_C\langle R \rangle - \mu_m  - \left(h + \sigma_C^2 \langle\chi\rangle \right) N + z_N \sqrt{\sigma^2_m + \sigma^2_h N^2 + \epsilon^2 \sigma_C^2 \langle R^2 \rangle } \right],\\
0 &= - \gamma^{-1}\epsilon \sigma_C^2 \langle \nu \rangle R^2- \left( \mu_D +\gamma^{-1}\mu_C \langle N \rangle \right) R + \mu_K  + z_R \sqrt{\sigma^2_K + \left( \sigma^2_D + \gamma^{-1}\sigma_C^2 \langle N^2 \rangle \right) R^2},
\end{aligned}
\label{eq:Supple-DMFT_results}
\end{equation}
where $ z_N \sqrt{\sigma^2_m + \sigma^2_h N^2 + \epsilon^2 \sigma_C^2 \langle R^2 \rangle }=\xi^N +\zeta^N N $ and $z_R \sqrt{\sigma^2_K + ( \sigma^2_D + \gamma^{-1}\sigma_C^2 \langle N^2 \rangle )R^2} =\zeta^R - \xi^R R$, and $z_N$ and $z_R$ are independent unit Gaussian random variables.
The results are the same as the cavity solution of Eq.~\eqref{eq:Supple-cavity_results}.
As $\epsilon$ is fixed to $1$ in this study, it is omitted in the following for the sake of simplicity.

\subsection{Abundance distributions for $\sigma_h=0$}
\label{supsubsec:Supple-II_C}
At the steady state, some consumers go extinct ($N = 0$) and the others survive with positive abundance $N = ( g_\text{eff} + z_N \sigma_g )/h_\text{eff}$, where $g_\text{eff}~( = \mu_C\langle R \rangle - \mu_m)$ is the effective growth rate, $h_\text{eff}~(= h+\sigma^2_C\langle\chi\rangle)$ is the effective intraspecific suppression coefficient, and $\sigma^2_g~(= \sigma^2_m + \sigma^2_C \langle R^2 \rangle)$ is the variance in the effective growth rate.
The resources have positive values for their abundance because the resources are supplied from outside with a constant input rate.
To simplify the effective dynamics in Eq.~\eqref{eq:Supple-cavity_results} and Eq.~\eqref{eq:Supple-DMFT_results}, we define the effective degradation rate $D_\text{eff} ~ (= \mu_D + \gamma^{-1} \mu_C \langle N \rangle )$ and variance $\sigma^2_R ~ (= \sigma^2_D + \gamma^{-1} \sigma^2_C \langle N^2 \rangle)$ in effective degradation rate.
Then, the abundances at the steady state in Eq.~\eqref{eq:Supple-cavity_results} and Eq.~\eqref{eq:Supple-DMFT_results} are
\begin{equation}
\label{eq:Supple-unit_random}
\begin{aligned}
z_N &= ( h_\text{eff} N - g_\text{eff})/\sigma_g \quad\text{for}~N>0,\\
z_R &= (\gamma^{-1}\sigma^2_C \langle \nu \rangle R^2 + D_\text{eff} R -\mu_K)/\sqrt{\sigma^2_K+\sigma^2_R R^2 }.
\end{aligned}
\end{equation}
From Eq.~\eqref{eq:Supple-unit_random}, we obtain abundance distributions of consumer species $P(N)$ and resources $P(R)$ using the change of variables, i.e., $P(N) = P(z_N) |dz_N/dN|$ and $P(R) = P(z_R) |dz_R/dR|$, where $P(z) = \frac{1}{\sqrt{2\pi}}e^{-z^2/2}$.
Here are the abundance distributions of consumer species and resources:
\begin{equation}
\label{eq:Supple-abundnace_distributions_GCRM}
\begin{aligned}
P(N) &= \frac{ h_\text{eff}}{\sqrt{2\pi} \sigma_g} \exp \left[- \frac{(h_\text{eff} N - g_\text{eff})^2}{2 \sigma^2_g } \right]\quad \text{for}~N>0,\\[15pt]
P(R) &= \frac{1}{\sqrt{2\pi}}\frac{\gamma^{-1}\sigma_R^2 \sigma_C^2 \langle\nu\rangle R^3 + (\sigma_R^2 \mu_K + 2\gamma^{-1}\sigma_K^2 \sigma_C^2 \langle\nu\rangle) R + \sigma_K^2 D_\text{eff}}{(\sigma^2_K + \sigma^2_R R^2 )^{3/2}} \exp \left[ - \frac{(\gamma^{-1}\sigma^2_C \langle\nu\rangle R^2 + D_\text{eff} R - \mu_K)^2}{2(\sigma^2_K + \sigma^2_R R^2 )} \right].
\end{aligned}
\end{equation}
Taking derivative on Eq.~\eqref{eq:Supple-DMFT_results} and using the relation in Eq.~\eqref{eq:Supple-unit_random}, we obtain the response functions,
\begin{equation}
\label{eq:Supple-responses}
\begin{aligned}
\nu(N) &= - \frac{\partial N}{\partial \mu_m} = \frac{1}{h_\text{eff}}=\frac{1}{h+\sigma_C^2 \langle \chi \rangle},\\[5pt]
\chi(R) &= - \frac{\partial R}{\partial \mu_D} = \frac{(\sigma^2_K + \sigma^2_R R^2)R}{\gamma^{-1}\sigma_R^2 \sigma_C^2 \langle\nu\rangle R^3 + (\sigma_R^2 \mu_K + 2\gamma^{-1}\sigma_K^2 \sigma_C^2 \langle\nu\rangle) R + \sigma_K^2 D_\text{eff}}.
\end{aligned}
\end{equation}
Note that to know the statistics of $N$, we should know the statistics of $R$ before, but information for $N$ is requisites to know the statistics of $R$.
It means that they depend on themselves and make self-consistent relations.

Now, we obtain the self-consistent equations for seven macroscopic quantities as follows:
\begin{equation}
\label{eq:Supple-cavity_final_results}
\begin{aligned}
\phi_S &= \int^\infty_{+0} dN\, P(N),\\
\langle N \rangle &= \int^\infty_{+0} dN\, N P(N),\\
\langle N^2 \rangle &= \int^\infty_{+0} dN\, N^2 P(N),\\
\langle \nu \rangle &= \int^\infty_{+0} dN\, \nu(N) P(N),\\
\langle R \rangle &= \int^\infty_{+0} dR\, R P(R),\\
\langle R^2 \rangle &= \int^\infty_{+0} dR\, R^2 P(R),\\
\langle \chi \rangle &=  \int^\infty_{+0} dR\, \chi(R) P(R).
\end{aligned}
\end{equation}

%For nonzero $\sigma_h$, the Jacobian term in $P(N)$ from Eq.~\eqref{eq:Supple-abundnace_distributions_GCRM} becomes zero at $N = -\sigma^2_g h_\text{eff} / \sigma^2_h g_\text{eff}$.
%This result implies that the consumer species are forbidden to have abundance value $N=-\sigma^2_g h_\text{eff}/\sigma^2_h g_\text{eff}$ because the value gives zero probability.
%The meaning of this forbidden value of $N$ remains a question.
%For zero $\sigma_h$, however, the forbidden value of $N$ disappears, because $P(N)$ in Eq.~\eqref{eq:Supple-abundnace_distributions_GCRM} becomes a simple truncated Gaussian distribution as
%\begin{equation}
%P(N) = \frac{ h_\text{eff} }{\sqrt{2\pi}\sigma_g}  \exp \left[- \frac{(h_\text{eff} N - %g_\text{eff})^2}{2 \sigma^2_g} \right]\quad \text{for}~N>0.
%\end{equation}

\begin{figure}[h]
\begin{center}
\includegraphics[width=1.00\linewidth]{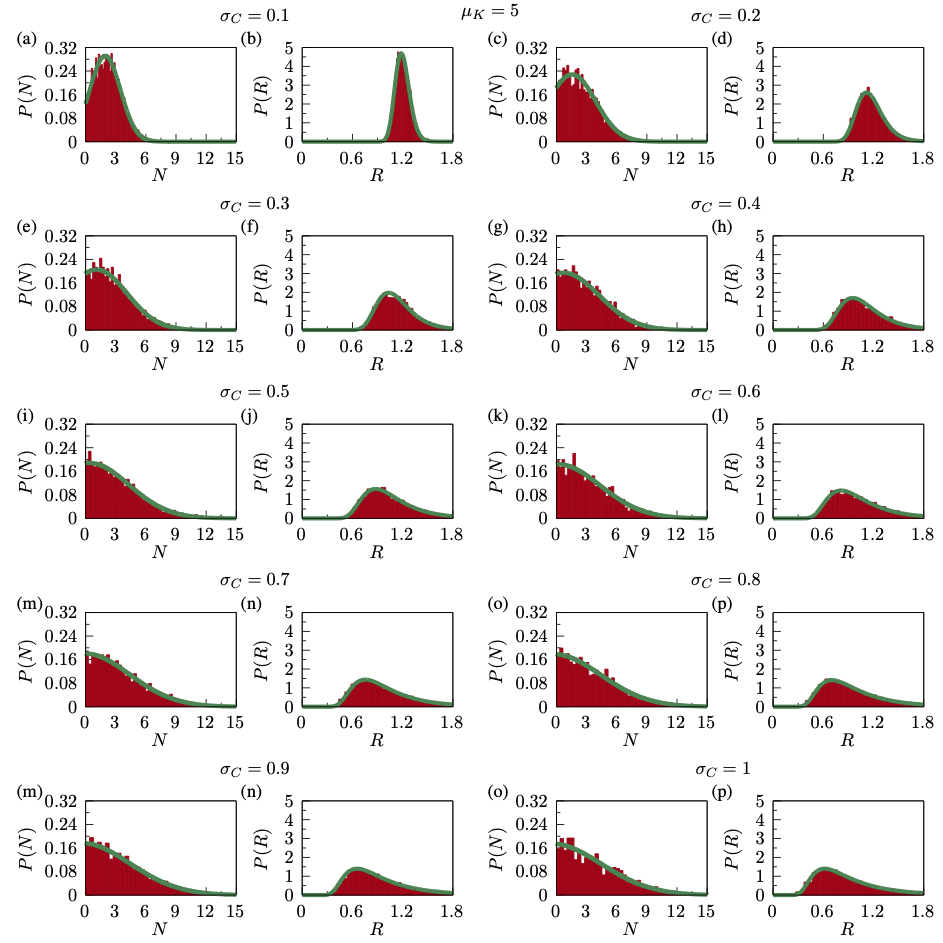}
\caption{
Consumer and resource abundance distributions with intraspecific suppression ($h=0$) in the resource-rich ($\mu_K=5$) environment for different $\sigma_C$.
}
\label{fig:Supple-Dist_with_rich}
\end{center}
\end{figure}

\begin{figure}[h]
\begin{center}
\includegraphics[width=1.00\linewidth]{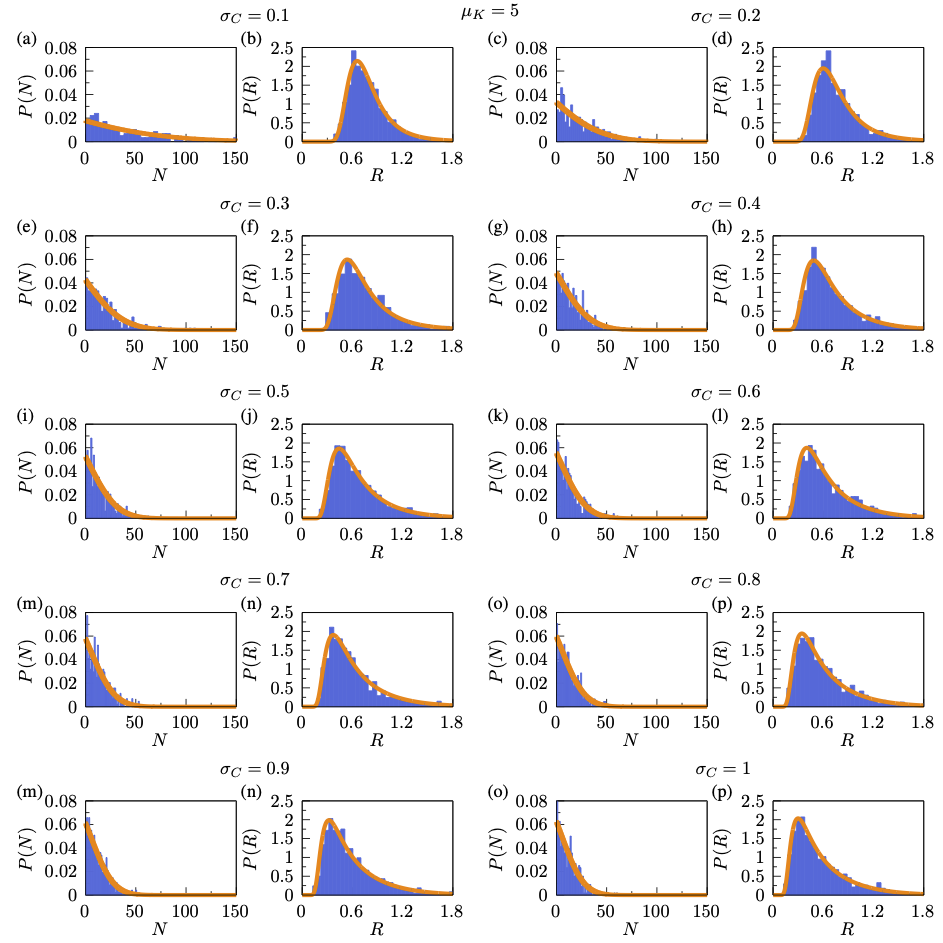}
\caption{
Consumer and resource abundance distributions without intraspecific suppression ($h=1/10$) in the resource-rich ($\mu_K=5$) environment for different $\sigma_C$.
}
\label{fig:Supple-Dist_without_rich}
\end{center}
\end{figure}

\begin{figure}[h]
\begin{center}
\includegraphics[width=1.00\linewidth]{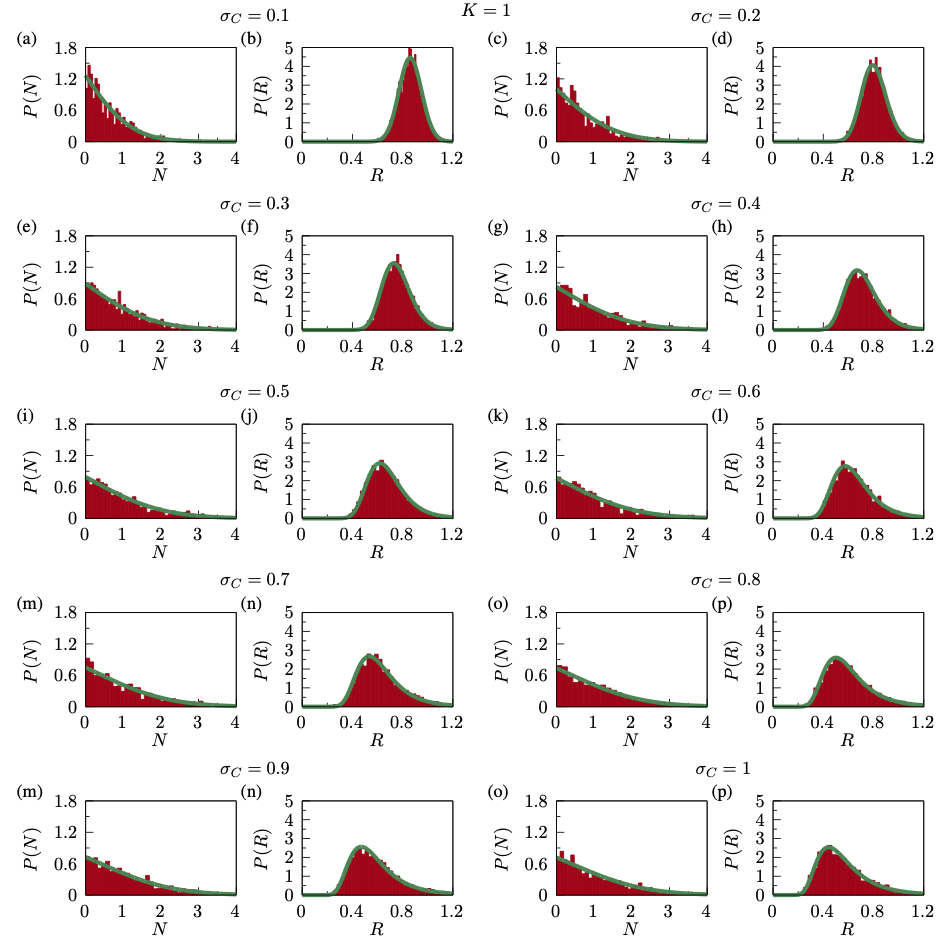}
\caption{
Consumer and resource abundance distributions with intraspecific suppression ($h=0$) in the resource-moderate ($\mu_K=1$) environment for different $\sigma_C$.
}
\label{fig:Supple-Dist_with_moderate}
\end{center}
\end{figure}

\begin{figure}[h]
\begin{center}
\includegraphics[width=1.00\linewidth]{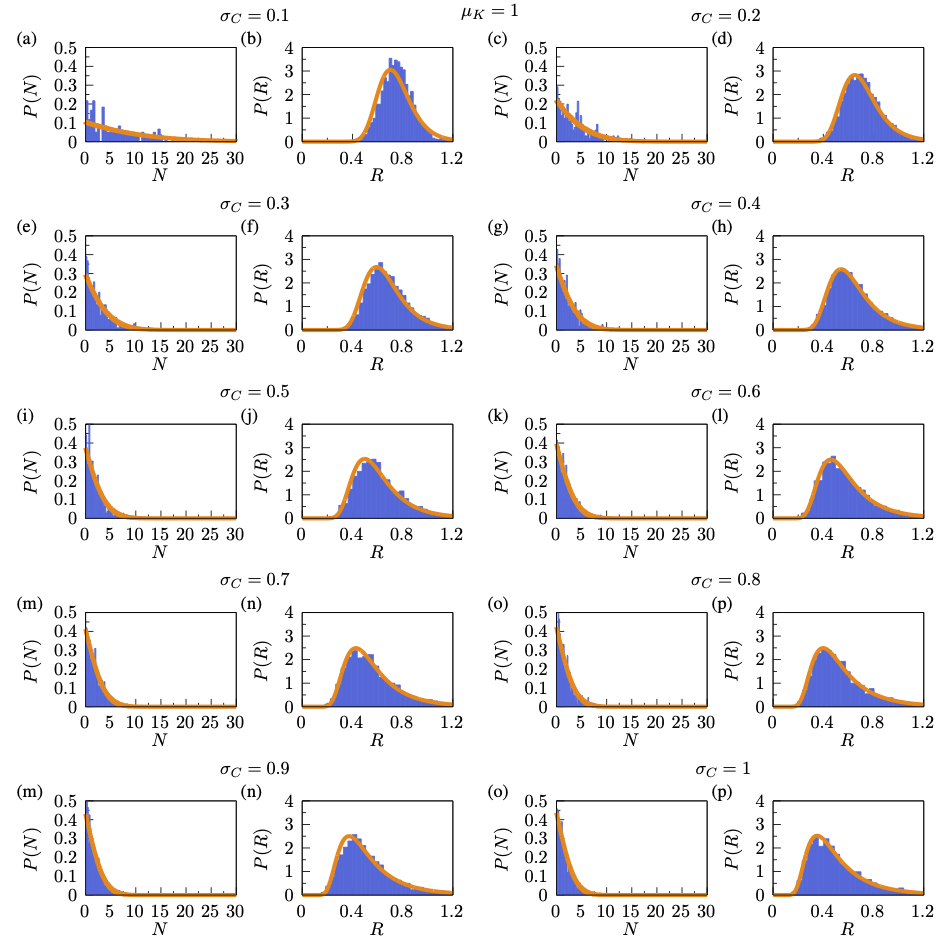}
\caption{
Consumer and resource abundance distributions without intraspecific suppression ($h=1/10$) in the resource-moderate ($\mu_K=1$) environment for different $\sigma_C$.
}
\label{fig:Supple-Dist_without_moderate}
\end{center}
\end{figure}

\begin{figure}[h]
\begin{center}
\includegraphics[width=1.00\linewidth]{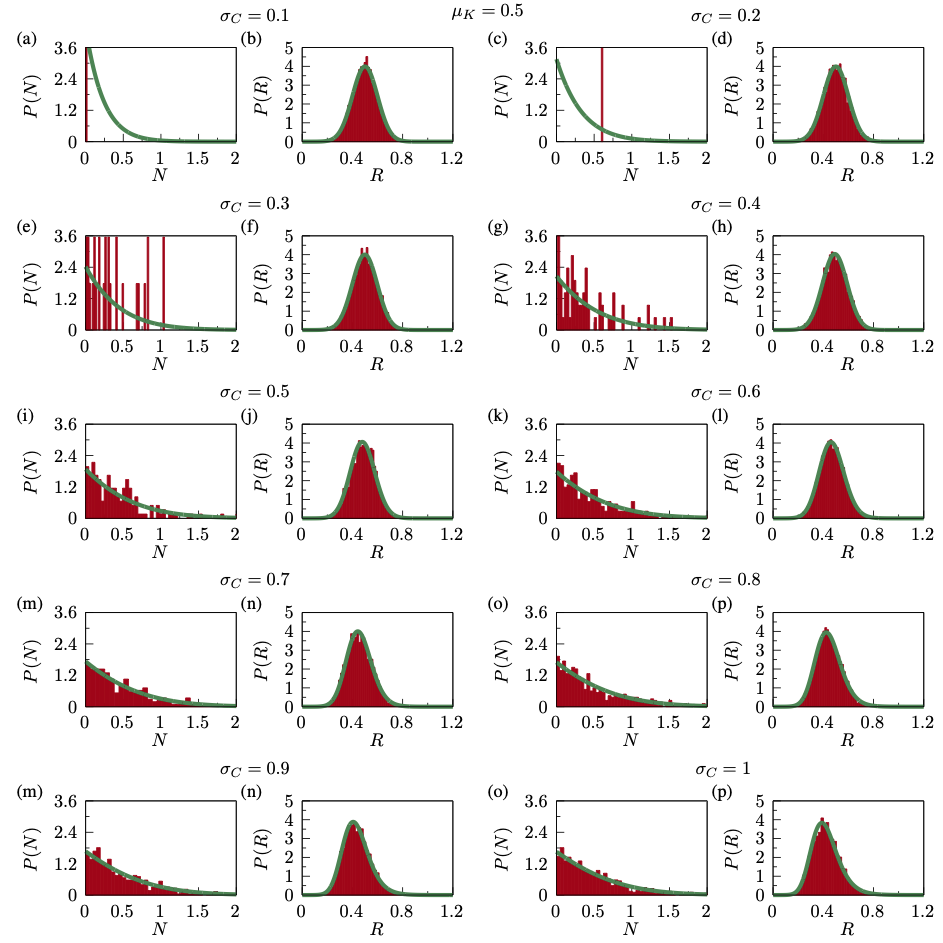}
\caption{
Consumer and resource abundance distributions with intraspecific suppression ($h=0$) in the resource-poor ($\mu_K=1/2$) environment for different $\sigma_C$.
For $\sigma_C\leq 0.3$, the surviving probability of consumers is too low to get enough data points to fit well the truncated Gaussian distribution for $N$.
}
\label{fig:Supple-Dist_with_poor}
\end{center}
\end{figure}

\begin{figure}[h]
\begin{center}
\includegraphics[width=1.00\linewidth]{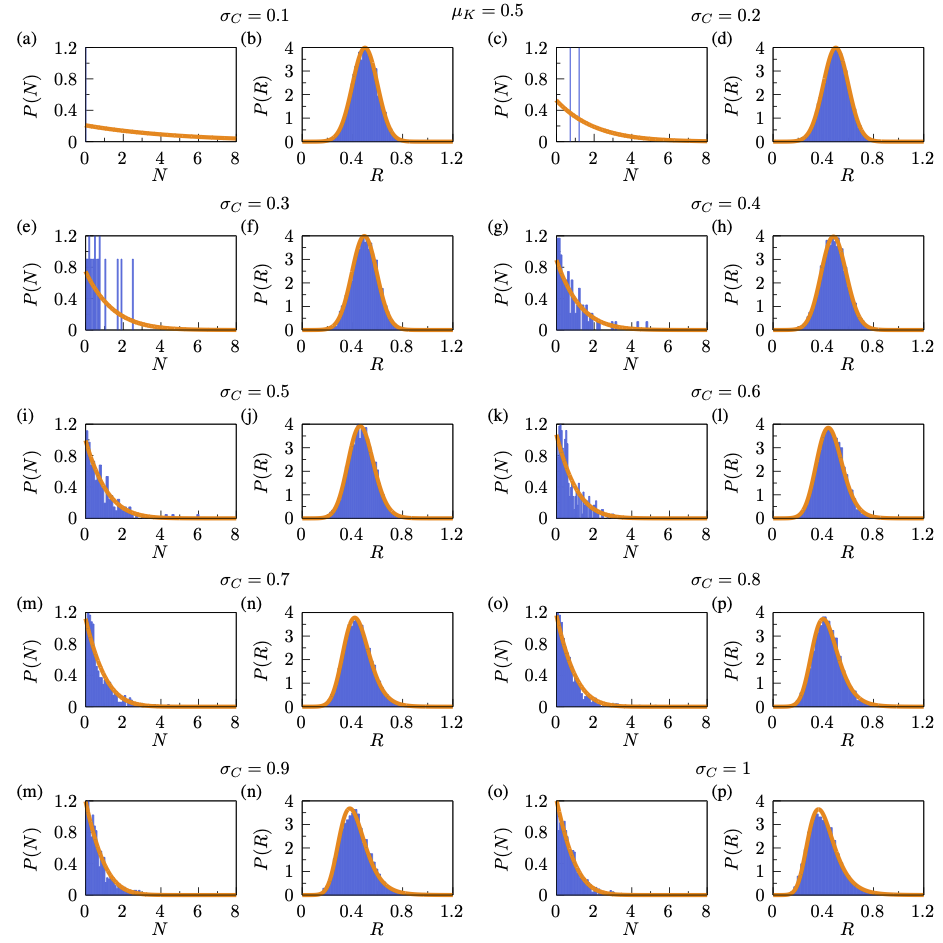}
\caption{
Consumer and resource abundance distributions without intraspecific suppression ($h=1/10$) in the resource-poor ($\mu_K=1/2$) environment for different $\sigma_C$.
For $\sigma_C\leq 0.4$, the number of data points is not sufficient to fit well the truncated Gaussian distribution for $N$.
}
\label{fig:Supple-Dist_without_poor}
\end{center}
\end{figure}

Figures~\ref{fig:Supple-Dist_with_rich}--\ref{fig:Supple-Dist_without_poor} show $P(N)$ and $P(R)$ with and without intraspecific suppression in three different environments for different $\sigma_C$.
In this case, the response function $\langle\nu\rangle$ becomes much simpler as $\langle\nu\rangle = \int^\infty_{+0} dN\, \nu(N)P(N) = \phi_S/h_\text{eff}$.

\begin{figure}[h]
\begin{center}
\includegraphics[width=0.56\linewidth]{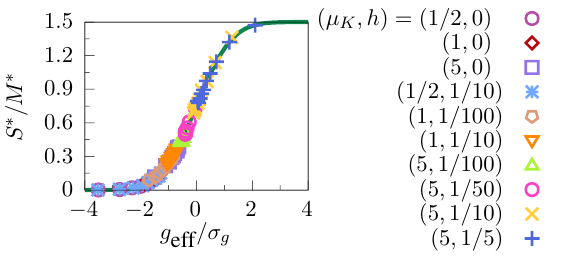}
\caption{
Relative diversity $S^*/M^*$ for different $(\mu_K,h)$ sets.
For each $(\mu_K,h)$, $\sigma_C$ varies in range $[0,1]$ with constant increment $d \sigma_C=0.1$.
The green solid line behind the symbols indicates $\gamma^{-1}/2 \left[ 1 + \text{erf} \left( g_\text{eff}/ \sqrt{2} \sigma_g \right) \right]$.
}
\label{fig:Supple-ErrorFtn}
\end{center}
\end{figure}

From our results, we obtain a relation between the relative diversity $S^*/M^*$ and other parameters as
\begin{equation}
\label{eq:Supple-SM}
S^*/M^* = \gamma^{-1}\phi_S = \gamma^{-1}h_\text{eff}\langle\nu\rangle = \gamma^{-1}(h+\sigma^2_C \langle\chi\rangle)\langle\nu\rangle.
\end{equation}
The $S^*/M^*$ also can be evaluated by integrating $P(N)$ from $+0$ to $\infty$ as
\begin{equation}
\label{eq:Supple-ErrorFtn}
S^*/M^* = \gamma^{-1}\phi_S = \gamma^{-1} \int_{+0}^\infty dN\, P(N) = \frac{\gamma^{-1}}{2}\left[ 1 + \text{erf} \left(\frac{g_\text{eff}}{\sqrt{2} \sigma_g} \right) \right],
\end{equation}
where $\text{erf}(x)$ indicates the error function.
We notice that $S^*/M^*$ is determined by the ratio between $g_\text{eff}$ and $\sigma_g$, and numerically validate Eq.~\eqref{eq:Supple-ErrorFtn} for different $K$ and $h$ varying $\sigma_C$ as displayed in Fig.~\ref{fig:Supple-ErrorFtn}.

%%%%%%%%%%%%%%%%%%%%%%%%%%%%%%%%%%%%%%%%%%%
\subsection{Bounds of the relative diversity}
\label{supsubsec:Supple-II_D}

In Eq.~\eqref{eq:Supple-SM}, the term $\gamma^{-1}\sigma^2_C \langle\chi\rangle \langle\nu\rangle$ increases in $\sigma_C$ starting from $0$ at $\sigma_C=0$, and thus $S^*/M^*$ has its minimum value at $\sigma_C=0$, and the maximum value as $\sigma_C \rightarrow \infty$. 
We approximately evaluate $\gamma^{-1}\sigma^2_C \langle\chi\rangle \langle\nu\rangle$ taking the limit of infinite $\sigma_C$ to determine the maximum value as
\begin{equation}
\begin{aligned}\label{eq:Supple-limit_half}
&\lim_{\sigma_C \to\infty} \gamma^{-1}\sigma^2_C \langle\chi\rangle \langle\nu\rangle = \lim_{\sigma_C \to\infty} \int^\infty_{+0} dR\, \gamma^{-1}\sigma^2_C \langle\nu\rangle \chi(R) P(R) \\[6pt]
&=\lim_{\sigma_C \to\infty}   \int^\infty_{+0} dR\, \frac{\gamma^{-1}\sigma^2_C \langle\nu\rangle R}{\sqrt{2\pi (\sigma^2_R R^2 + \sigma^2_K)}}\exp \left[ -\frac{ ( \gamma^{-1} \sigma^2_C \langle\nu\rangle R^2 + D_\text{eff} R - \mu_K )^2 }{2(\sigma^2_R R^2 + \sigma^2_K)} \right]\\[6pt]
&=\lim_{\sigma_C \to\infty}  \int^\infty_{+0} dR\, \frac{\gamma^{-1}\sigma^2_C \langle\nu\rangle R}{\sqrt{2\pi [ (\gamma^{-1}\sigma^2_C\langle N^2\rangle + \sigma^2_D) R^2 + \sigma^2_K ]}}\exp \left[ -\frac{ ( \gamma^{-1} \sigma^2_C \langle\nu\rangle R^2 + D_\text{eff} R - \mu_K )^2 }{2[(\gamma^{-1}\sigma^2_C\langle N^2\rangle + \sigma^2_D) R^2 + \sigma^2_K]} \right] \\[6pt]
&\approx \lim_{\sigma_C \to\infty}  \int^\infty_{+0} dR\, \frac{\gamma^{-1/2}\sigma_C \langle\nu\rangle}{\sqrt{2\pi \langle N^2 \rangle }} \left( 1 -  \frac{\sigma^2_D R^2 + \sigma^2_K}{2 \gamma^{-1}\sigma^2_C \langle N^2 \rangle R^2} \right) \exp \left[ -\frac{\gamma^{-1}\sigma^2_C\langle\nu\rangle^2 R^2 }{2\langle N^2 \rangle} \left( 1 + \frac{D_\text{eff}R - \mu_K}{\gamma^{-1}\sigma^2_C\langle\nu\rangle R} \right)^2 \left( 1 - \frac{\sigma^2_D R^2 + \sigma^2_K}{\gamma^{-1}\sigma_C^2\langle N^2 \rangle R^2}\right) \right]\\[6pt]
&\approx \lim_{\sigma_C \to\infty}   \int^\infty_{+0} dR\, \frac{\gamma^{-1/2}\sigma_C \langle\nu\rangle}{\sqrt{2\pi \langle N^2 \rangle }}\exp \left[ -\frac{ \gamma^{-1} \sigma^2_C \langle\nu\rangle^2 R^2 }{2 \langle N^2 \rangle} \right]
= \frac{1}{2}.
\end{aligned}
\end{equation}
In the last approximation step, we used the fact that $\sigma^2_C\nu$ and $\sigma^2_C \langle N^2 \rangle$ diverge as $\sigma_C \rightarrow \infty$.
Finally, we determine the bounds of $S^*/M^*$ as follows:
\begin{equation}
\label{eq:Supple-bound_ext_GCRM}
\gamma^{-1}h\langle\nu\rangle \leq S^*/M^* < \gamma^{-1}h\langle\nu\rangle + 1/2.
\end{equation}

%%%%%%%%%%%%%%%%%%%%%%%%%%%%%%%%%%%%%%%%%%%
\subsection{Nonzero $\sigma_h$ case}
\label{supsubsec:Supple-II_E}
When the intraspecific suppression is not uniform, i.e., $\sigma_h\neq 0$, $z_N$ in Eq.~\eqref{eq:Supple-unit_random} changes as
\begin{equation}
\label{eq:Supple-unit_random_nonidentical}
\begin{aligned}
z_N &= ( h_\text{eff} N - g_\text{eff})/\sqrt{\sigma^2_g + \sigma^2_h N^2} \quad\text{for}~N>0.
\end{aligned}
\end{equation}
In the same manner as written in Sec.~\ref{supsubsec:Supple-II_C}, we obtain consumer abundance distribution $P(N)$ and $\nu(N)$ from Eq.~\eqref{eq:Supple-unit_random_nonidentical} as
\begin{equation}
\label{eq:Supple-abundnace_distributions_GCRM_nonidentical}
\begin{aligned}
P(N) &= \frac{1}{\sqrt{2\pi}} \left| \frac{ \sigma^2_g h_\text{eff} + \sigma^2_h g_\text{eff}  N }{( \sigma^2_g + \sigma^2_h N^2)^{3/2}} \right| \exp \left[- \frac{(h_\text{eff} N - g_\text{eff})^2}{2 ( \sigma^2_g + \sigma^2_h N^2)} \right]\quad \text{for}~N>0,\\
\nu(N) &= - \frac{\partial N}{\partial \mu_m} = \frac{ \sigma^2_g + \sigma^2_h N^2}{  \sigma^2_g h_\text{eff} +\sigma^2_h g_\text{eff} N}.
\end{aligned}
\end{equation}
The resource abundance distribution $P(R)$ and the resource response function $\chi(R)$ remain as given in Eqs.~\eqref{eq:Supple-abundnace_distributions_GCRM} and \eqref{eq:Supple-responses}.
For nonzero $\sigma_h$, the Jacobian term in $P(N)$ from Eq.~\eqref{eq:Supple-abundnace_distributions_GCRM_nonidentical} becomes zero at a specific consumer abundance, $N = -\sigma^2_g h_\text{eff} / \sigma^2_h g_\text{eff}$.
This result implies that consumers cannot attain the abundance value $N=-\sigma^2_g h_\text{eff}/\sigma^2_h g_\text{eff}$, which appears to be a forbidden state.
The underlying mechanism and implication of this forbidden value needs further investigation.

\begin{figure}[h]
\begin{center}
\includegraphics[width=1.00\linewidth]{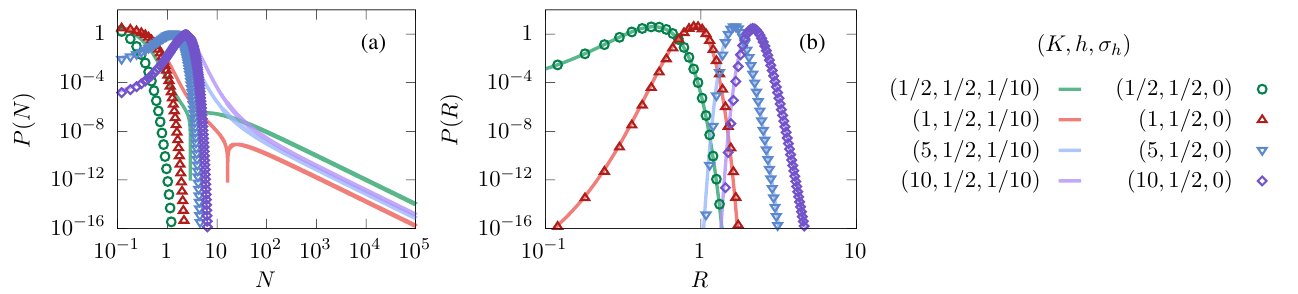}
\caption{
Abundance distributions for identical and nonidentical intraspecific suppression.
(a) The consumer abundance distribution $P(N)$ shows fat tail for nonidentical suppression.
(b) The resource abundance distribution $P(R)$ shows log-normal like distribution for sufficient resource input.
}
\label{fig:Supple-Ps_nonidentical}
\end{center}
\end{figure}

Figure~\ref{fig:Supple-Ps_nonidentical} presents the abundance distributions for the identical ($\sigma_h=0)$ and non-identical ($\sigma_h\neq0$) suppression cases.
Interestingly, although we consider the Gaussian interactions in the model described in Eq.~\eqref{eq:model_ext}, the consumer abundance distribution $P(N)$ exhibits a fat tail for non-identical suppression.
Moreover, when the resource input is sufficiently large, the resource abundance distribution $P(R)$ shows a lognormal-like shape in both identical and non-identical suppression.

It is noteworthy that $\sigma_h$ must be sufficiently small; otherwise, some values of $h_i$ may become negative, causing divergence in the consumer abundance $N_i$.
This occurs because the intraspecific suppression term $-h_i N_i^2$ may become positive, destabilizing the system.

\begin{figure}[h]
\centering\includegraphics[width=0.35\linewidth]{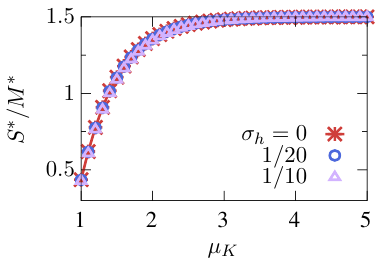}
\caption{
Relative diversity $S^*/M^*$ for three different deviations $\sigma_h$ of intraspecific suppression.
The absolute differences of $S^*/M^*$ between zero $\sigma_h$ case and nonzero $\sigma_h$ cases are less than $2\times 10^{-2}$.
The mean intraspecific suppression coefficient is fixed to $h=1/2$.
}\label{fig:Supple-Hetero}
\end{figure}

In terms of the relative diversity $S^*/M^*$, for nonzero $\sigma_h$ case exhibits qualitatively the same behavior as zero $\sigma_h$ case as shown in Fig.~\ref{fig:Supple-Hetero}.

%%%%%%%%%%%%%%%%%%%%%%%%%%%%%%%%%%%%%%%%%%%
\subsection{Intraspecific suppression}
\label{supsubsec:Supple-II_F}
In this section, we discuss two different mechanisms of intraspecific suppression; the first is direct intraspecific competition from random encounter, and the other is growth suppression by pathogen spreading.

\subsubsection{From random encounter}
In the reaction scheme, the dynamics can be written as 
\begin{equation}
\begin{aligned}
I_i + L_\alpha &\xrightarrow{C_{i\alpha}} I_i + I_i,\\
I_i &\xrightarrow{m_i} \varnothing,\\
I_i + I_i &\xrightarrow{h_i} I_i,
\end{aligned}
\end{equation}
where $I_i$  and $L_\alpha$ represents an individual of species $i$ and resource $\alpha$, respectively.
Let us define $X_i$ as the number of individuals of species $i$ and $Y_\alpha$ as the number of units of resource $\alpha$.
In a well-mixed system, the probability that an individual of species $i$ encounters a unit of resource $\alpha$ is proportional to $Y_{\alpha}/\Omega$ where $\Omega$ is a contextual parameter that can be interpreted as volume or carrying capacity.
In the same way, the probability of encountering another individual of species $i$ is proportional to $(X_i-1)/\Omega$.
As the number of individuals of species $i$ is $X_i$, the chance of such an event is proportional to $X_i$, and thus, the abundance dynamics of consumer species $i$ can be written by 
\begin{equation}
\begin{aligned}
\dot{X}_i &= X_i \left(\sum_\alpha C_{i\alpha} Y_\alpha/\Omega - m_i \right) - h_i{X_i (X_i-1)/\Omega}.\\
\end{aligned}
\end{equation}
Once we scale $X_i$ and $Y_\alpha$ in terms of $K$ calling them abundances, $X_i/\Omega \equiv N_i$ and $Y_\alpha/\Omega \equiv R_\alpha$, we obtain
\begin{equation}
\begin{aligned}
\dot{N}_i &= N_i \left(\sum_\alpha C_{i\alpha} R_\alpha - m_i \right) - h_i{N_i (N_i-1/\Omega)},\\
& = N_i \left( \sum_\alpha C_{i\alpha} R_\alpha - m_i - h_i N_i + \frac{h_i}{\Omega} \right),\\
& \approx N_i \left( \sum_\alpha C_{i\alpha} R_\alpha - m_i - h_i N_i \right),
\end{aligned}
\end{equation}
where the last approximation is valid when $1\ll \Omega$.

\subsubsection{From pathogen spreading}
\begin{CJK*}{UTF8}{}
\CJKfamily{mj}

While the previous study~\cite{MCRM_Monod} thoroughly explained how intraspecific suppression arises from pathogen spread, we will revisit this derivation here.
We expand GCRM to include pathogen abundance dynamics as follows:
\begin{equation}
\begin{aligned}
\dot{X}_\text{ㄱ} &= X_\text{ㄱ} \left( \sum_i J_{\text{ㄱ}i}N_i - u_\text{ㄱ} X_\text{ㄱ}\right) ,\\
\dot{N}_i &= N_i \left(\sum_\alpha C_{i\alpha} R_\alpha - m_i - \sum_\text{ㄱ} J_{\text{ㄱ}i} X_\text{ㄱ}\right),\\
\dot{R}_\alpha &= K_\alpha - \left( D_\alpha + \sum_i C_{i\alpha} N_i \right),
\end{aligned}
\end{equation}
where $X_\text{ㄱ}$ represents the abundance of pathogen $\text{ㄱ}$, and the pathogen grows with consumption rate $J_{\text{ㄱ}i}$.
The growth suppression of consumer species by pathogen is proportional to abundance of consumer species.
As a result, more abundant species experience higher suppression by pathogen, which aligns with `Kill-the-Winner' hypothesis.
The carrying capacity of the pathogen $\text{ㄱ}$ is determined by $u_\text{ㄱ}$.

We assume that pathogen dynamics occur on a much shorter timescale (fast dynamics assumption).
Consequently, we can express $X_\text{ㄱ}=\frac{1}{u_\text{ㄱ}}\sum_i J_{\text{ㄱ}i}N_i$.
With the fast pathogen dynamics assumption, the consumer abundance dynamics turns out
\begin{equation}
\begin{aligned}
\dot{N}_i &= N_i \left(\sum_\alpha C_{i\alpha} R_\alpha - m_i - \sum_\text{ㄱ} J_{\text{ㄱ}i} X_\text{ㄱ}\right)\\
&= N_i \left(\sum_\alpha C_{i\alpha} R_\alpha - m_i - \sum_j \sum_\text{ㄱ} \frac{J_{\text{ㄱ}i} J_{\text{ㄱ}j}}{u_\text{ㄱ}} N_j\right)\\
&= N_i \left(\sum_\alpha C_{i\alpha} R_\alpha - m_i - \sum_j \sum_\text{ㄱ} J'_{\text{ㄱ}i} J'_{\text{ㄱ}j} N_j\right).
\end{aligned}
\label{eq:pathogen}
\end{equation}
For the sake of simplicity, we rescale the consumption rate $J$ of pathogen with $u$, i.e., $J'_{\text{ㄱ}i} = J_{\text{ㄱ}i}/\sqrt{u_\text{ㄱ}}$.
Like the consumption rate $C_{i\alpha}$, one can treat $J'_{\text{ㄱ}i}$ as a random variable randomly drawn from $\mathcal{N}(J/S_P, \sigma^2_J/S_P)$, i.e., $J'_{\text{ㄱ}i} = J/S_P + \sigma_J/\sqrt{S_P} z^J_{\text{ㄱ}i}$, where $S_P (\gg 1)$ is the number of pathogen species, and $z^J_{\text{ㄱ}i}$ is the unit Gaussian random variable.

The last term in Eq.~\eqref{eq:pathogen} describes the growth suppression by pathogen spreading among consumer species.
The term is rewritten as
\begin{equation}
\begin{aligned}
&\sum_j \sum_\text{ㄱ} J'_{\text{ㄱ}i}J'_{\text{ㄱ}j} N_i N_j = \sum_j \sum_\text{ㄱ} (J/S_P + \sigma_J/\sqrt{S_P} z^J_{\text{ㄱ}i})(J/S_P + \sigma_J/\sqrt{S_P} z^J_{\text{ㄱ}j}) N_i N_j\\
&=\sum_j \sum_\text{ㄱ} \left[ \frac{J^2}{S^2_P} +  \frac{J \sigma_J}{S_P\sqrt{S_P}}\left( z^J_{\text{ㄱ}i} + z^J_{\text{ㄱ}j} \right) + \frac{\sigma^2_J}{S_P} z^J_{\text{ㄱ}i}z^J_{\text{ㄱ}j} \right] N_i N_j\\
&\approx \frac{J^2}{S_P} \sum_j  N_i N_j + \sigma^2_J N^2_i = J^2 \frac{S}{S_P}N_i \langle N \rangle + \sigma^2_J N^2_i.
\end{aligned}
\end{equation}
We utilized the property of independent unit Gaussian random variable $z^J$ above.
In this study, we ignore the term $J^2 \frac{S}{S_P}N_i \langle N \rangle$.
Finally, we obtain the consumer abundance dynamics as shown in this study:
\begin{equation}
\begin{aligned}
\dot{N}_i &= N_i \left(\sum_\alpha C_{i\alpha} R_\alpha - m_i - \sigma^2_J N_i \right) = N_i \left(\sum_\alpha C_{i\alpha} R_\alpha - m_i - h N_i \right).
\end{aligned}
\end{equation}
\end{CJK*}
% One can obtain the same expression of intraspecific suppression in Eq.~(1) in the main text by assuming that the pathogen spreads only within the same species.

%%%%%%%%%%%%%%%%%%%%%%%%%%%%%%%%%%%%%%%%%%
\clearpage
\label{supsec:Supple-III}
\section{Numerical Methods and Results}

\subsection{Numerical methods}
\label{subsec:numer_methods}
We numerically integrate Eq.~\eqref{eq:model_ext} for $50$ independent realizations of the model parameters with random initial conditions using the Runge-Kutta-Fehlberg (RKF) method~\cite{RK45}.
For each realization, we use $S=75$ consumers and $M=50$ resources.
The RKF method changes the step size $\Delta t$ adaptively by comparing the integration results in the order of $\mathcal{O}(\Delta t^4)$ and $\mathcal{O}(\Delta t^5)$.
In the numerical integration, we regard a species goes extinct when its abundance has a lower value than the extinction threshold $10^{-9}$.

We also obtain the solutions of self-consistent equations using the optimization method, Nelder-Mead algorithm~\cite{Nelder_Mead}.
Our optimization function is basically designed as the sum of difference squares between the given values of seven macroscopic quantities, $\phi_S$, $\langle N \rangle$, $\langle N^2 \rangle$, $\langle\nu\rangle$, $\langle R \rangle$, $\langle R^2 \rangle$, and $\langle\chi\rangle$.
The corresponding values are obtained by integrating the abundance distributions $P(N)$ and $P(R)$, as described in Eqs.~\eqref{eq:Supple-abundnace_distributions_GCRM} and \eqref{eq:Supple-cavity_final_results}. 
The algorithm aims to minimize this optimization function by seeking appropriate values for the quantities.
We set the tolerance parameter for the optimization to $10^{-30}$.
All codes are publicly available on GitHub~\cite{my_git}.
%

%%%%%%%%%%%%%%%%%%%%%%%%%%%%%%%%%%%%%%%%%
\subsection{Numerical results}

\begin{figure}[h]
\begin{center}
\includegraphics[width=1.00\linewidth]{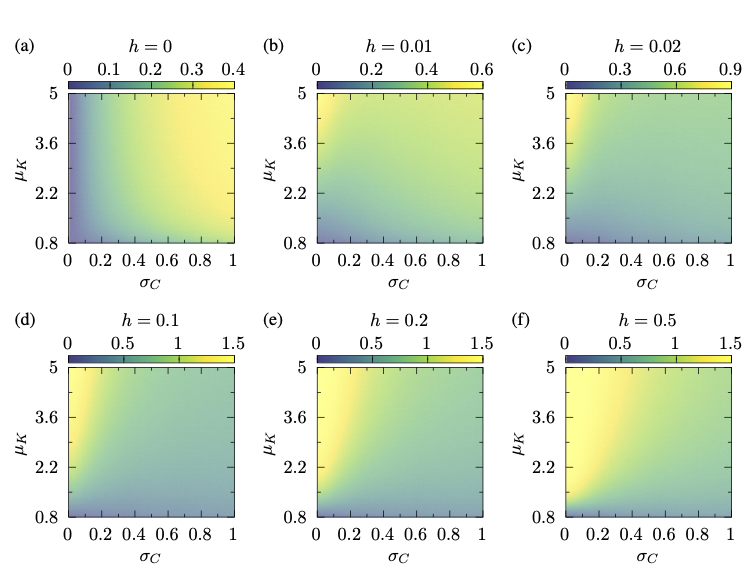}
\caption{
Relative diversity $S^*/M^*$ in $(\sigma_C,\mu_K)$-plane in different environments of $h=0,~0.01,~0.02,~0.1,~0.2$, and $0.5$.
}
\label{fig:Supple-SM_K_sigC}
\end{center}
\end{figure}
\begin{figure}[h]
\begin{center}
\includegraphics[width=1.00\linewidth]{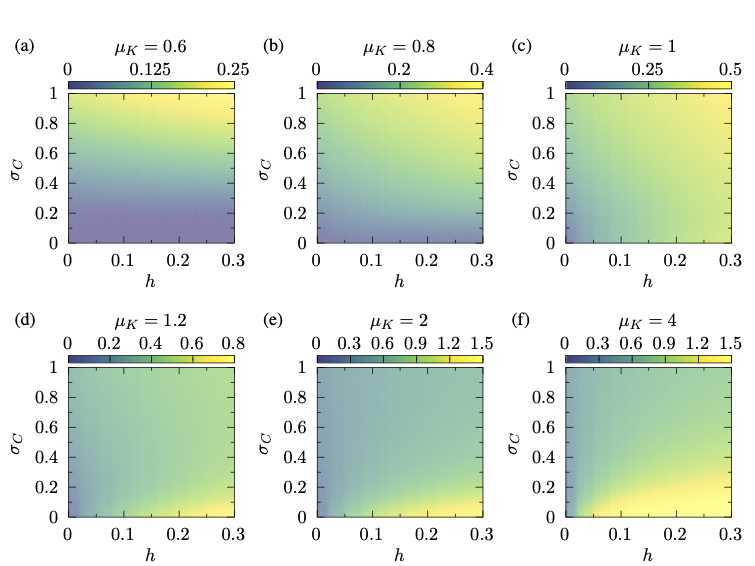}
\caption{
Relative diversity $S^*/M^*$ in $(h,\sigma_C)$-plane in different environments of $\mu_K=0.6,~0.8,~1,~1.2,~2$, and $4$.
}
\label{fig:Supple-SM_forKs}
\end{center}
\end{figure}
\begin{figure}[h]
\begin{center}
\includegraphics[width=1.00\linewidth]{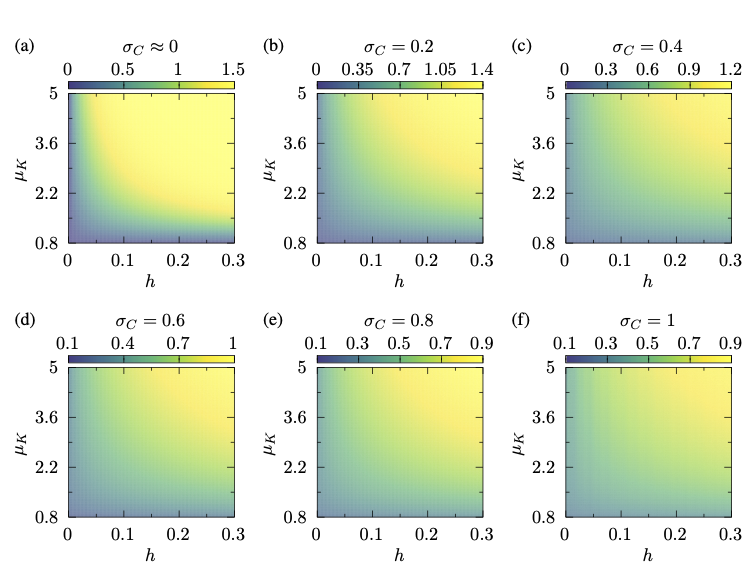}
\caption{
Relative diversity $S^*/M^*$ in $(h,\mu_K)$-plane for six different consumption rate deviations of $\sigma_C=0.004,~0.2,~0.4,~0.6,~0.8$, and $1$.
}
\label{fig:Supple-SM_Kh}
\end{center}
\end{figure}
\begin{figure}[h]
\begin{center}
\includegraphics[width=1.00\linewidth]{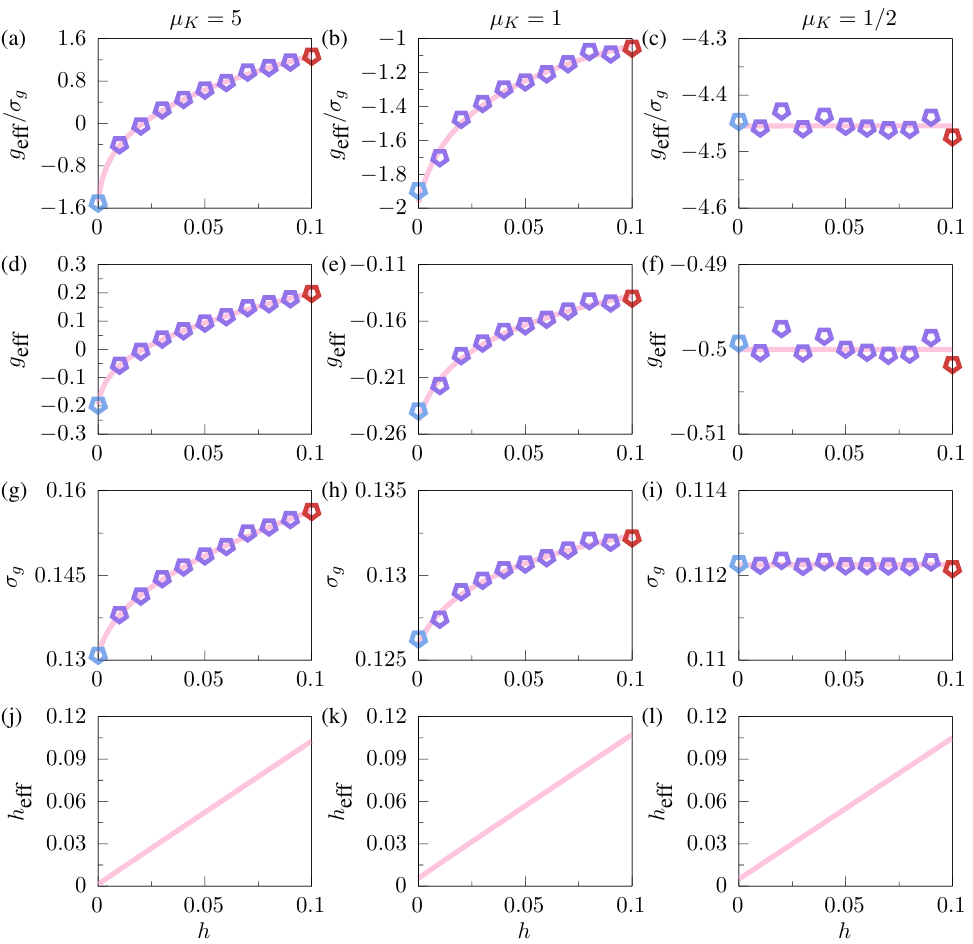}
\caption{
[(a), (b), and (c)] Ratio between effective growth rate $g_\text{eff}$ and deviation $\sigma_g$ in growth rate, [(d), (e) and (f)] $g_\text{eff}$, [(g), (h) and (i)] $\sigma_g$, and [(j), (k), and (l)] effective intraspecific suppression coefficient $h_\text{eff}$ versus intraspecific suppression coefficient $h$ for $\sigma_C=1/10$ in [(a), (d), (g), and (j)] resource-rich ($\mu_K=5$), [(b), (e), (h), and (k)] resource-moderate ($\mu_K=1$), and [(c), (f), (i), and (l)] resource-poor ($\mu_K=1/2$) environments.
The pink solid line denotes the analytic result.
The symbols are obtained from averaging over $50$ independent realizations.
In the resource-poor environment, the number of surviving consumer species is too small, thus, the symbols quite deviate from analytic results.
}
\label{fig:Supple-effective_parameters_h}
\end{center}
\end{figure}
\begin{figure}[h]
\begin{center}
\includegraphics[width=1.00\linewidth]{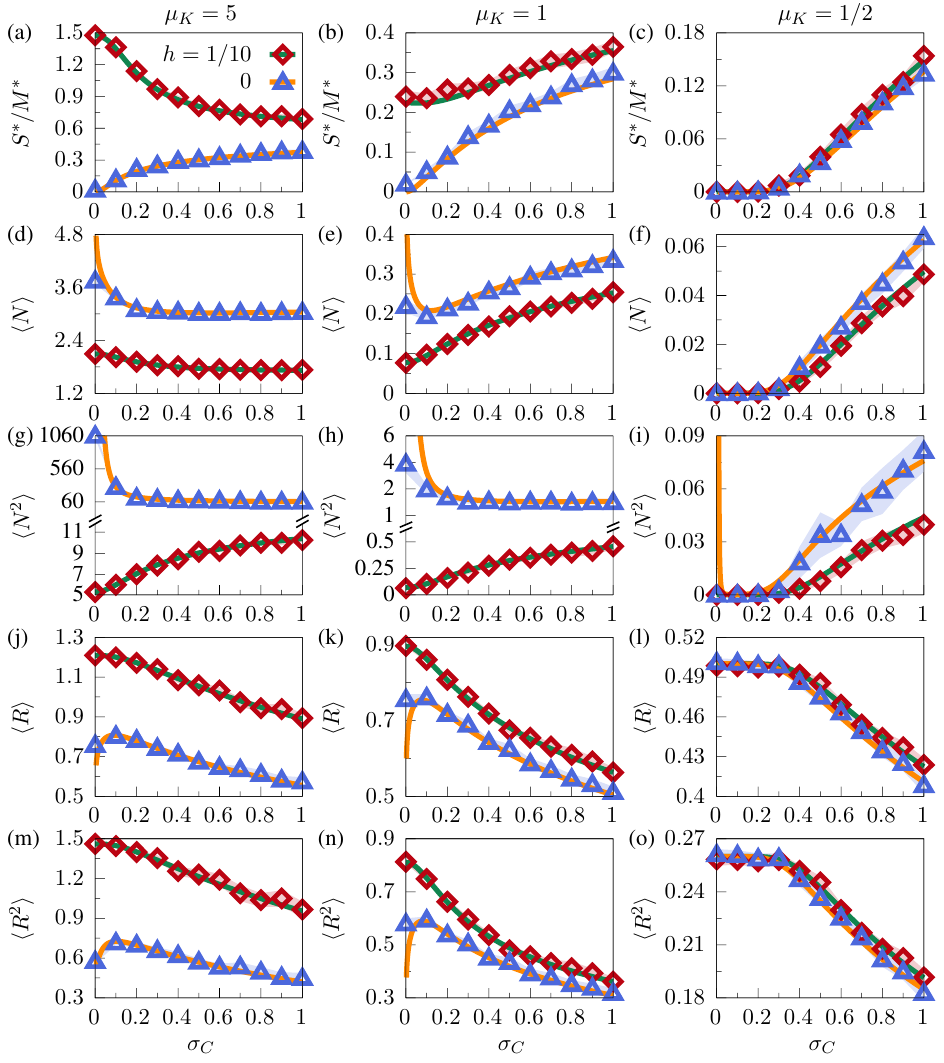}
\caption{
[(a), (b), and (c)] Relative diversity $S^*/M^*$ of consumers 
, [(d), (e), and (f)] mean consumer abundance $\langle N \rangle$, [(g), (h), and (i)] mean squared consumer abundance $\langle N^2 \rangle$, [(j), (k), and (l)] mean resource abundance $\langle R \rangle$, and [(m), (n), and (o)] mean squared resource abundance $\langle R^2 \rangle$ versus consumption rate deviation $\sigma_C$ in [(a), (d), (g), (j), and (m)] resource-rich $(\mu_K=5)$, [(b), (e), (h), (k), and (n)] resource-moderate ($\mu_K=1$), and [(c), (f), (i), (l), and (o)] resource-poor ($\mu_K=1/2$) environments.
The solid lines are obtained from solving self-consistent equations, and the symbols are obtained from numerical simulations averaging over $50$ different realizations.
The colored shades indicate twice the standard error.
}
\label{fig:Supple-GCRM_Results}
\end{center}
\end{figure}
\begin{figure}[h]
\begin{center}
\includegraphics[width=1.00\linewidth]{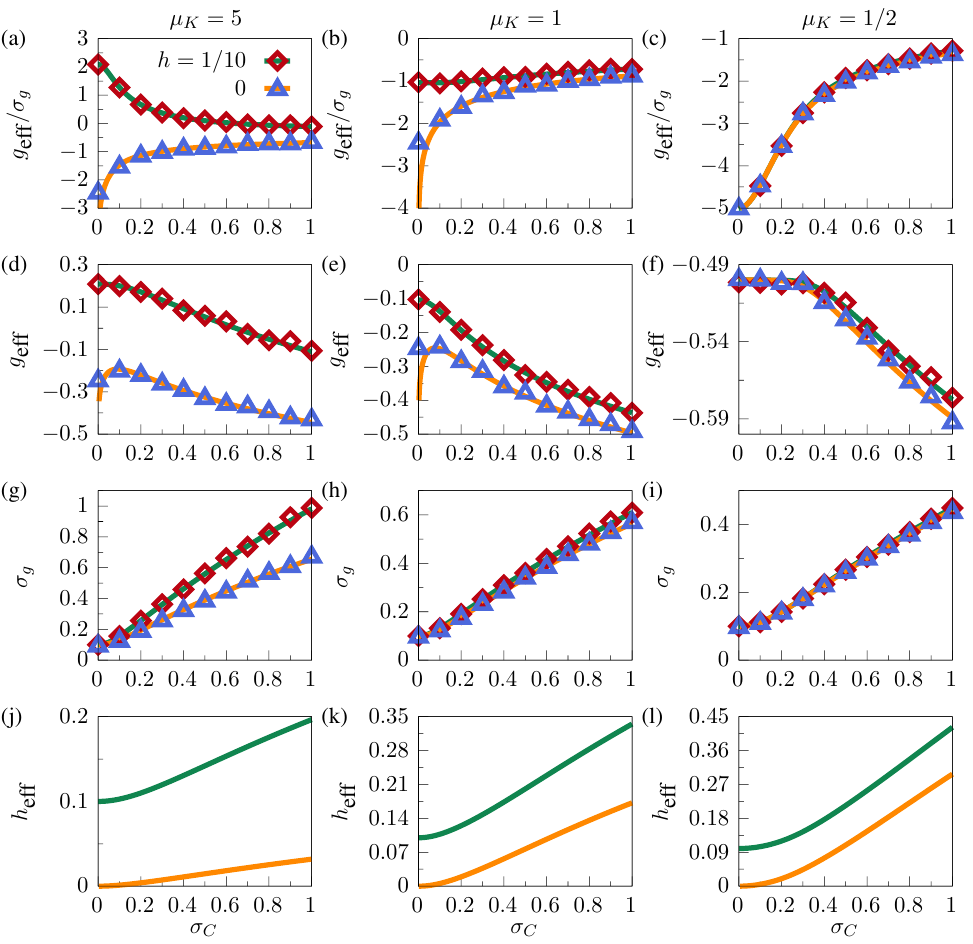}
\caption{
[(a), (b), and (c)] Ratio between effective growth rate $g_\text{eff}$ and deviation $\sigma_g$ in growth rate, [(d), (e) and (f)] $g_\text{eff}$, [(g), (h) and (i)] $\sigma_g$, and
[(j), (k), and (l)] effective intraspecific suppression coefficient $h_\text{eff}$ versus consumption rate deviation $\sigma_C$ for without ($h=0$), and with ($h=1/10$) intraspecific suppression in [(a), (d), (g), and (j)] resource-rich ($\mu_K=5$), [(b), (e), (h), and (k)] resource-moderate ($\mu_K=1$), and [(c), (f), (i), and (l)] resource-poor ($\mu_K=1/2$) environments.
The green and yellow solid lines indicate the analytic results with and without intraspecific suppression, respectively.
The symbols are obtained from numerical simulations averaging over $50$ independent realizations.
}
\label{fig:Supple-effective_parameters}
\end{center}
\end{figure}
%

%Rank Distribution
%
\begin{figure}[h]
\begin{center}
\includegraphics[width=0.70\linewidth]{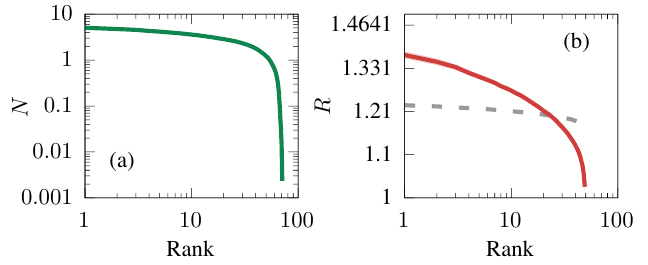}
\caption{
Rank distributions of (a) consumer abundance $N$ and (b) resource abundance $R$ for $(K,h) = (5,1/10)$ are shown on a log-log scale.
(a) The rank distribution of $N$ exhibits the characteristic shape of a Gaussian distribution.
(b) In contrast, the rank distribution of $R$ shows a fat tail, as commonly observed in lognormal or power-law distributions.
%The shaded areas represent one standard error.
The gray dotted line in (b) represents the rank distribution of a Gaussian distribution with the same mean and standard deviation as $P(R)$.
}
\label{fig:Supple-RankDist}
\end{center}
\end{figure}

%
%Responses
%
\begin{figure}[h]
\begin{center}
\includegraphics[width=0.70\linewidth]{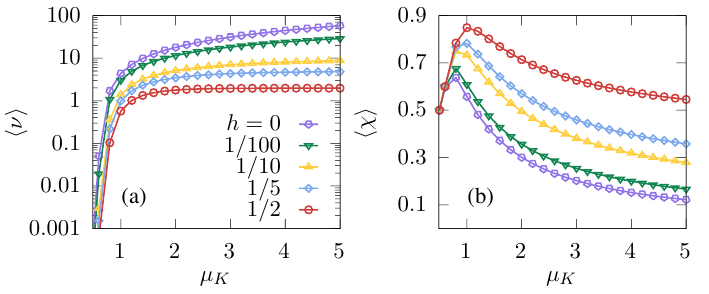}
\caption{
(a) Consumer response function $\langle\nu\rangle$ and (b) Resource response function $\langle\chi\rangle$ against resource input rate $\mu_K$. 
}
\label{fig:Supple-Responses}
\end{center}
\end{figure}

\clearpage
\label{supsec:Supple-IV}
\section{Biotic Resources}
\begin{figure}[h]
\begin{center}
\includegraphics[width=0.90\linewidth]{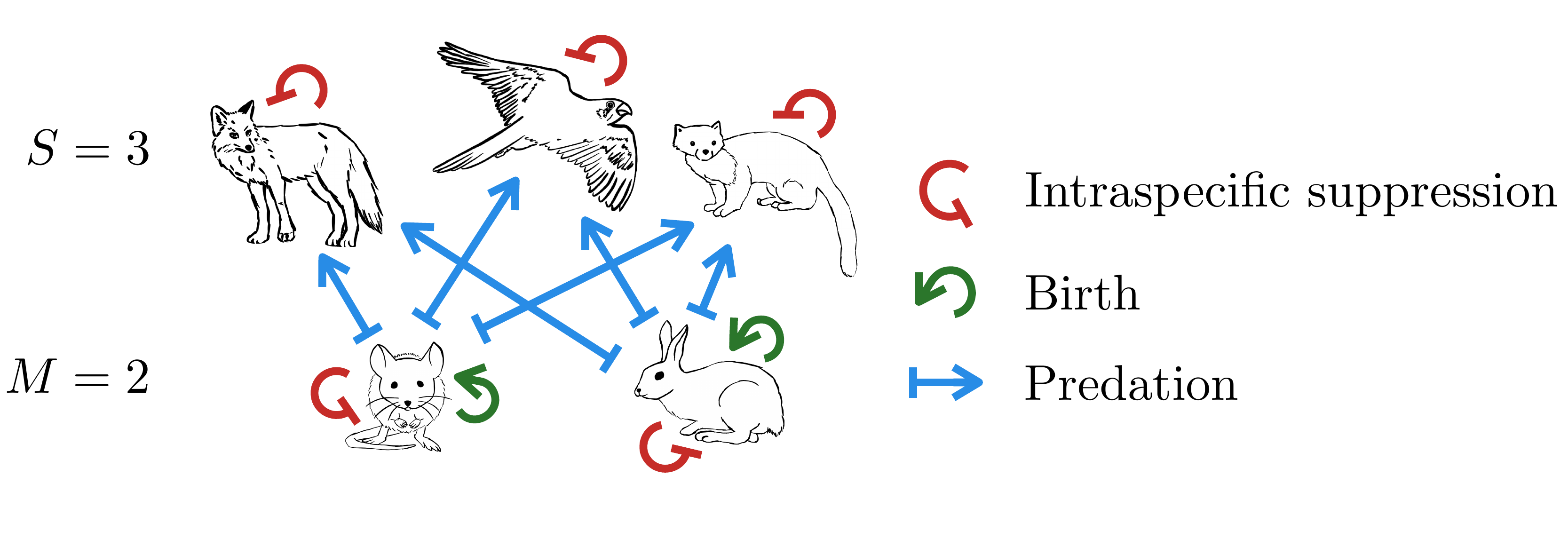}
\caption{
Schematic figure of an ecological system that consists of three predators ($S=3$) and two biotic prey species ($M=2$).
Different from the abiotic resources, prey species give birth and grow by themselves.
}
\label{fig:Supple-PP_Schemematic}
\end{center}
\end{figure}

The GCRM for biotic resources without intraspecific suppression has been well studied~\cite{Supple-GCRM_Cavity}.
In this section, the model with intraspecific suppression is dealt with.
The GCRM for biotic resources with intraspecific suppression is written as follows:
\begin{equation}
\label{eq:Supple-model_PP}
\begin{aligned}
\dot{N}_i &= N_i \left( \sum_\alpha C_{i\alpha} R_\alpha -m_i -h_i N_i \right),\quad &i=1,2,\cdots,S, \\
\dot{R}_\alpha &= R_\alpha \left( r_\alpha - d_\alpha R_\alpha - \sum_i N_i C_{i\alpha} \right), \quad &\alpha=1,2,\cdots,M,
\end{aligned}
\end{equation}
where $r_\alpha$ and $d_\alpha$ denote the reproduction rate and the intraspecific suppression coefficient of biotic resource $\alpha$, respectively.

Following the procedures in Sec.~II, we get the result of
\begin{equation}
\label{eq:Supple-PP_simple}
z_R = (r_\text{eff} - d_\text{eff}R)/\sqrt{\sigma^2_R + \sigma^2_d R^2}\quad \text{for}\ R>0,
\end{equation}
where $r_\text{eff} (= \mu_r - \gamma^{-1}\mu_C\langle N \rangle)$ is the effective growth rate of biotic resources, $d_\text{eff}(= d + \gamma^{-1}\sigma^2_C \langle\nu\rangle)$ is the effective death rate of the resources, and $\sigma^2_R~(= \sigma^2_r + \gamma^{-1}\sigma^2_C\langle N^2 \rangle)$ is the variance in the effective growth rate of the resources.
The results for the consumer species are the same as described in Sec.~II.

Using the change of variables in the probability distribution, we obtain the abundance distribution for biotic resources as
\begin{equation}
\label{eq:Supple-PP_dist_R}
P(R) = \frac{1}{\sqrt{2\pi}} \left| \frac{\sigma^2_d r_\text{eff} R + \sigma^2_R d_\text{eff}}{(\sigma^2_d R^2 + \sigma^2_R)^{3/2}} \right| \exp\left[ -\frac{(d_\text{eff} R - r_\text{eff})^2}{2(\sigma^2_d R^2 + \sigma^2_R)} \right]\quad \text{for}\ R>0,
\end{equation}
and the response function as
$\chi(R) = \partial R/\partial r = (\sigma^2_R + \sigma^2_d R^2) / (\sigma^2_R d_\text{eff} + \sigma^2_d r_\text{eff} R)$.

Note that for this type of resource, the resources can also go extinct, which is different from the externally supplied abiotic resources.
Thus, the surviving probability $\phi_M$ of resources should be considered, and it is calculated by $\phi_M=\int^\infty_{+0} dR\, P(R)$.

\begin{figure}[h]
\begin{center}
\includegraphics[width=0.65\linewidth]{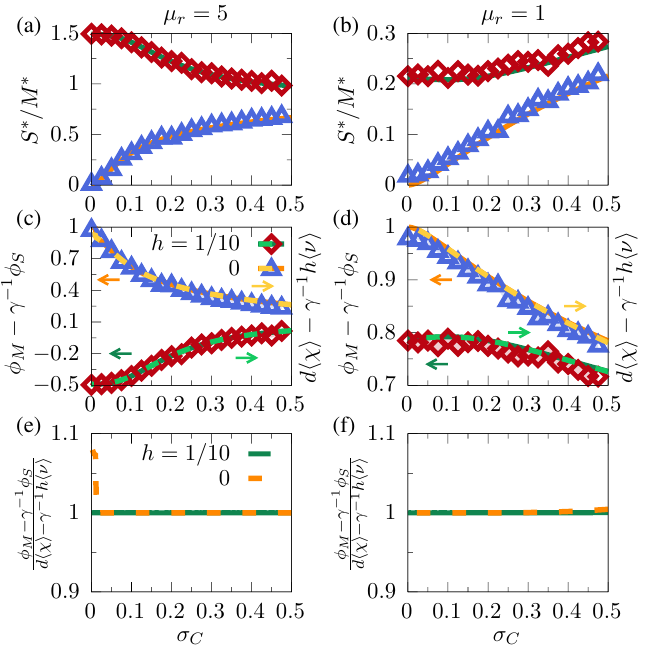}
\caption{
[(a) and (b)] The relative diversity $S^*/M^*$ for (a) highly ($\mu_r=5$) and (b) moderately ($\mu_r=1$) reproductive prey without ($h=0$) and with ($h=1/10$) intraspecific suppression for consumers.
[(c), (d), (e), and (f)] The relation we obtain for biotic resources for [(c) and (e)] highly and [(d) and (f)] moderately reproductive prey without and with intraspecific suppression for consumers.
[(e) and (f)] We validate our results by dividing the left-hand side with the right-hand side of Eq.~\eqref{eq:Supple-equality_self_PP}.
The symbols are obtained by averaging over $50$ independent realizations.
In (c) and (d), the orange (yellow) and green (chartreuse) solid (dotted) lines denote the left-hand (right-hand) side without and with intraspecific suppression, respectively.
Red and blue symbols are obtained by averaging over $50$ independent realizations.
In (e) and (f), the green solid line and orange dotted line indicate the ratio between left- and right-hand sides with and without intraspecific suppression, respectively.
}
\label{fig:Supple-relation_PP}
\end{center}
\end{figure}

When we ignore the disorder in intraspecific suppression for resources ($\sigma_d=0$), we obtain the truncated Gaussian resource abundance distribution like consumer abundance as
\begin{equation}
\label{eq:Supple-PP_distributions}
P(R) = \frac{ d_\text{eff}}{\sqrt{2\pi} \sigma_R} \exp\left[ -\frac{(d_\text{eff} R - r_\text{eff})^2}{2 \sigma^2_R} \right]\quad\text{for}\ R>0.
\end{equation}
In this situation, the resource response function also becomes simpler as $\chi(R) = 1/d_\text{eff}$, and this result gives $\langle\chi\rangle = \int_{+0}^\infty dR\, \chi(R)P(R)=\phi_M/d_\text{eff}$.
Utilizing the relation between $\phi_M$ and $\langle\chi\rangle$, and $\phi_S$ and $\langle\nu\rangle$, we obtain 
\begin{equation}
\label{eq:Supple-equality_self_PP}
\phi_M-\gamma^{-1}\phi_S =  d \langle\chi\rangle - \gamma^{-1}h\langle\nu\rangle, 
\end{equation}
where $d$ is the mean value of $d_\alpha$.
As $\phi_M-\gamma^{-1}\phi_S=(M^*-S^*)/M$, when coexisting consumers $S^*$ is larger than the number of available prey $M^*$, $\phi_M-\gamma^{-1}\phi_S$ is negative.
Thus by plotting $\phi_M-\gamma^{-1}\phi_S$ in Fig.~\ref{fig:Supple-relation_PP}, we investigate whether the intraspecific suppression enhances the relative diversity for biotic resources beyond the CEP bound $1$.
After a little bit more mathematics from Eq.~\eqref{eq:Supple-equality_self_PP}, we obtain
\begin{equation}
\label{eq:Supple-SM_PP}
S^*/M^* = \frac{\gamma^{-1}(h+\sigma_C^2\langle\chi\rangle)\langle\nu\rangle}{(d+\gamma^{-1}\sigma_C^2\langle\nu\rangle)\langle\chi\rangle}.
\end{equation}
This relation is also depicted in Figs.~\ref{fig:Supple-relation_PP}(a) and (b).
From Eq.~\eqref{eq:Supple-SM_PP}, we can identify the values of $S^*/M^*$ in two limiting cases, $\sigma_C = 0$ and $\sigma_C \rightarrow \infty$, as
$\lim_{\sigma_C=0} S^*/M^* = \gamma^{-1} h\langle\nu\rangle/d\langle\chi\rangle$ and $\lim_{\sigma_C\rightarrow \infty} S^*/M^* = 1$.

\begin{figure}[h]
\begin{center}
\includegraphics[width=0.65\linewidth]{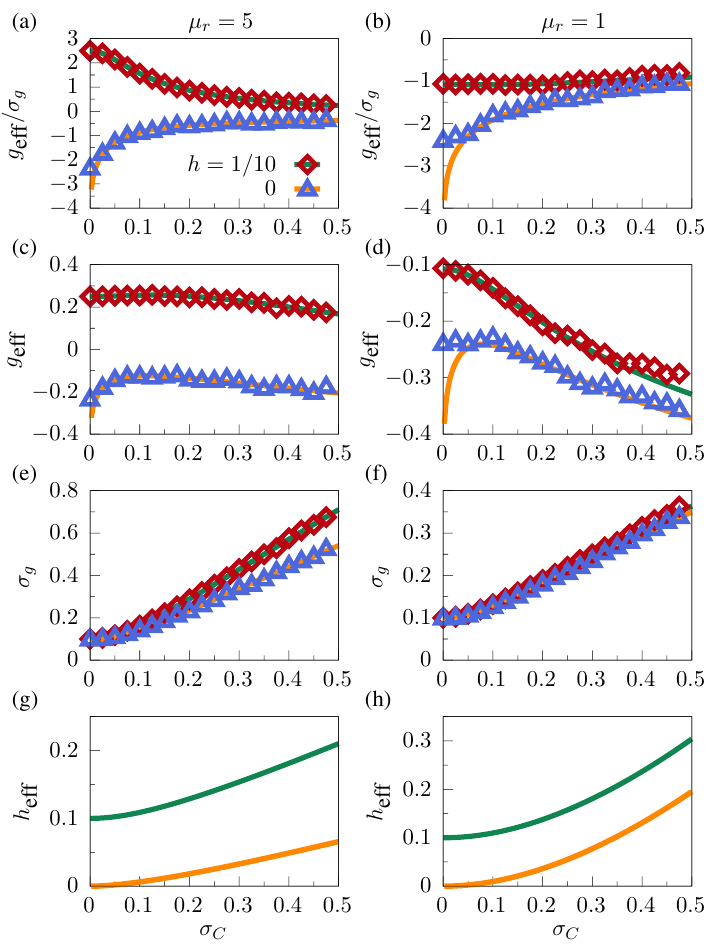}
\caption{
[(a) and (b)] The ratio between the effective growth rate $g_\text{eff}$ of consumers and the deviation $\sigma_g$ in the growth rate, [(c) and (d)] $g_\text{eff}$, [(e) and (f)] $\sigma_g$, and [(g) and (h)] the effective intraspecific suppression coefficient $h_\text{eff}$ with ($h=1/10$) and without ($h=0$) intraspecific suppression for [(a), (c), (e), and (g)] highly ($\mu_r=5$) and [(b), (d), (f), and (h)] moderately ($\mu_r=1/2$) reproductive prey.
}
\label{fig:Supple-SpeciesPacking_PP}
\end{center}
\end{figure}

For highly reproductive prey ($\mu_r=5$), the number of coexisting consumers is larger than the number of available prey in the system for $\sigma_C < 0.45$ with intraspecific suppression ($h=1/10$) [Fig.~\ref{fig:Supple-relation_PP}(a)].
The effective parameters of consumer species $g_\text{eff}$, $\sigma_g$, and $h_\text{eff}$ are displayed in Fig.~\ref{fig:Supple-SpeciesPacking_PP}.
We fix the parameters $\gamma=2/3$, $\mu_C=1$, $\mu_m=1$, $d=1$, $\sigma_m=1/10$, $\sigma_d=0$, $\sigma_r=1/10$, and $\sigma_h=0$ throughout this section.
For numerical simulations, we fix $S=75$ and $M=50$.

\clearpage
\begin{figure}[h]
\begin{center}
\includegraphics[width=1.00\linewidth]{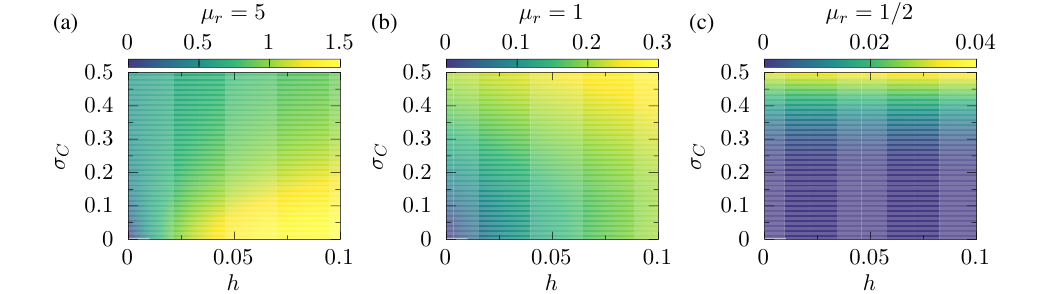}
\caption{
Relative diversity $S^*/M^*$ in $(h,\sigma_C)$-plane for two different reproduction rates of (a) $\mu_r=5$, (b) $\mu_r=1$, and (c) $\mu_r=1/2$.
The relative diversity shows qualitatively the same behavior as that in the externally supplied abiotic resource.
}
\label{fig:Supple-sigC_h_PP}
\end{center}
\end{figure}

\begin{figure}[h]
\begin{center}
\includegraphics[width=0.90\linewidth]{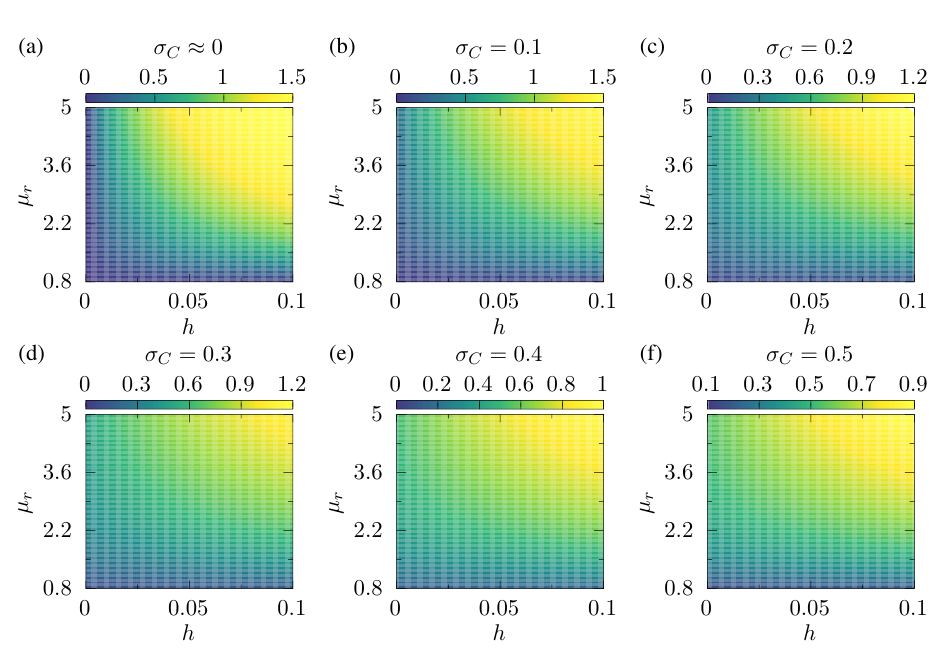}
\caption{
Relative diversity $S^*/M^*$ in $(h,\mu_r)$-plane for six different consumption rate deviations of $\sigma_C=0.004,~0.1,~0.2,~0.3,~0.4$, and $0.5$.
}
\label{fig:Supple-SM_Kh_PP}
\end{center}
\end{figure}

\begin{figure}[h]
\begin{center}
\includegraphics[width=1.00\linewidth]{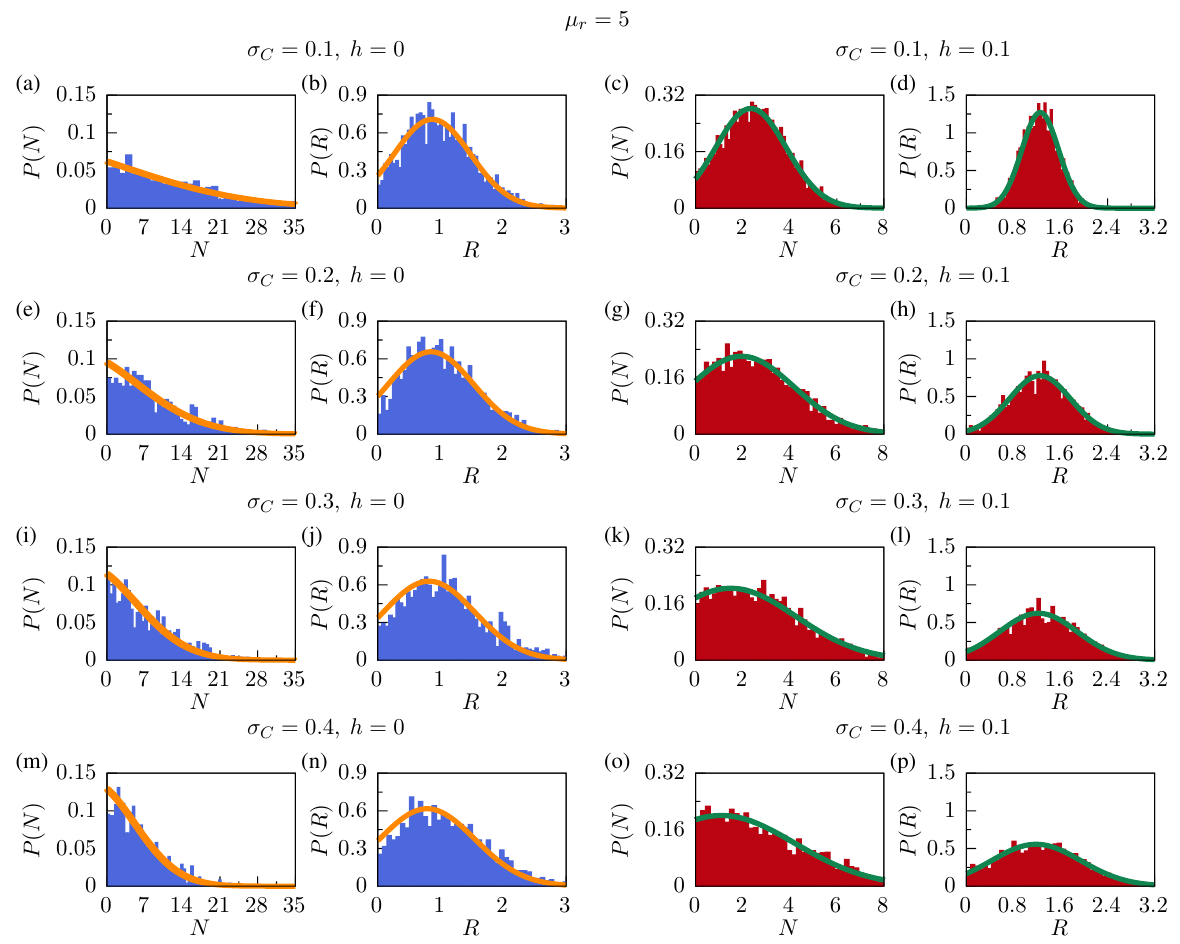}
\caption{
Distributions without $(h=0)$ and with ($h=1/10$) intraspecific suppression of consumers for highly ($\mu_r=5$) reproductive prey.
}
\label{fig:Supple-Dist_rich_PP}
\end{center}
\end{figure}

\begin{figure}[h]
\begin{center}
\includegraphics[width=1.00\linewidth]{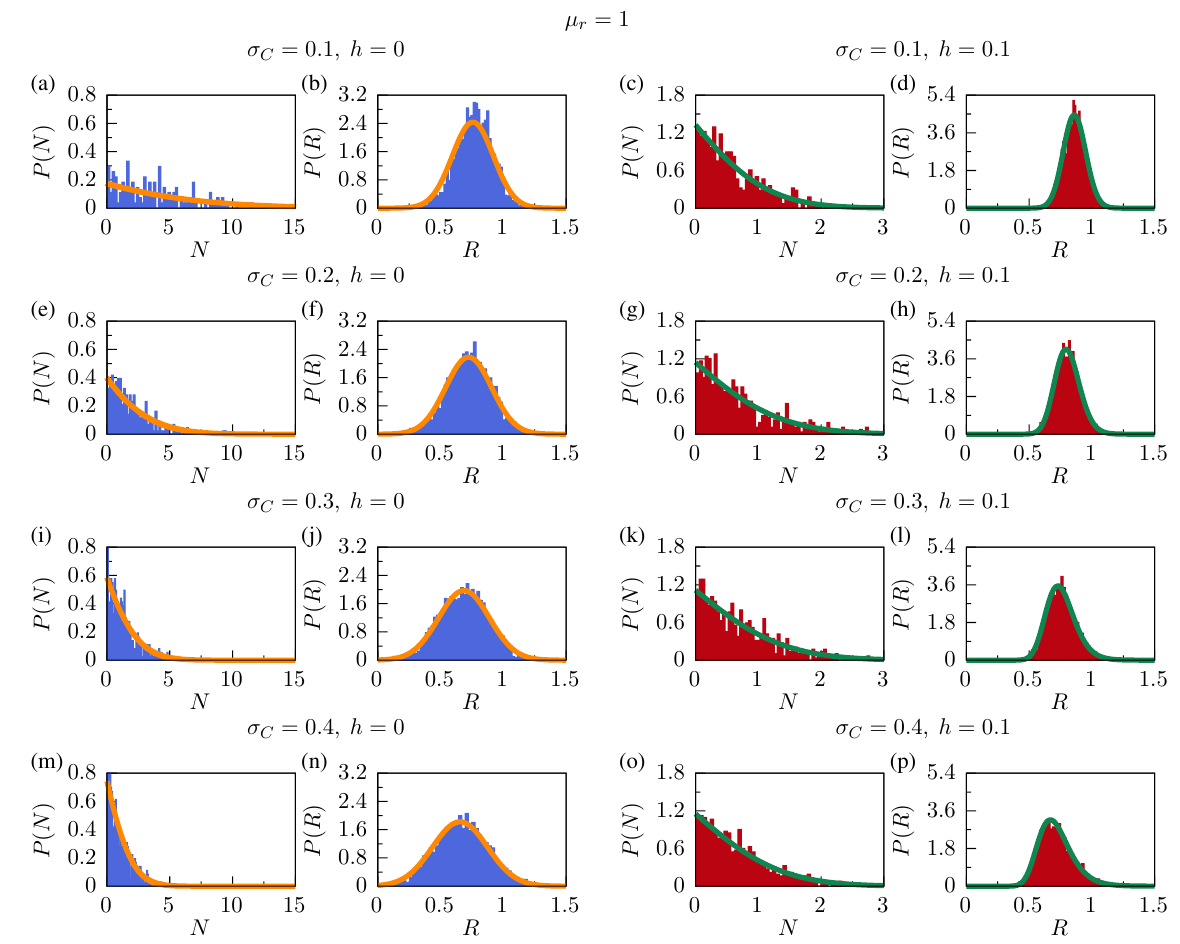}
\caption{
Distributions without $(h=0)$ and with ($h=1/10$) intraspecific suppression of consumers for moderately ($\mu_r=1$) reproductive prey.
}
\label{fig:Supple-Dist_moderate_PP}
\end{center}
\end{figure}

\clearpage
\label{supsec:Supple-V}
\section{Three-Trophic Level Systems}
\begin{CJK*}{UTF8}{}
\CJKfamily{mj}

In this Section, we extend the model to describe three-trophic level systems, which consist of consumers, producers, and externally supplied abiotic resources, and obtain the abundance distributions for two consumers and externally supplied abiotic resource utilizing the method described in Sec.~II.
The ecological systems of three-trophic level can be described by the following equations:
\begin{equation}
\label{eq:Supple-model_Three}
\begin{aligned}
\dot{N}_i &=N_i \left( \sum_\alpha C_{i\alpha} R_\alpha -m_i -h_i N_i \right), \quad &i=1,2,\cdots,S, \\[6pt]
\dot{R}_\alpha &= R_\alpha \left( r_\alpha + \sum_{\text{ㄱ}}G_{\alpha\text{ㄱ}} A_\text{ㄱ} - d_\alpha R_\alpha - \sum_i N_i C_{i\alpha} \right), \quad &\alpha=1,2,\cdots,M,\\[6pt]
\dot{A}_{\text{ㄱ}} &= K_\text{ㄱ} - \left(E_\text{ㄱ} + \sum_\alpha R_\alpha G_{\alpha\text{ㄱ}} \right)A_\text{ㄱ}, \quad &\text{ㄱ}=1,2,\cdots,L,
\end{aligned}
\end{equation}
where $N_i$, $R_\alpha$, and $A_\text{ㄱ}$ indicate the abundance of consumer $i$, producer $\alpha$, and resource $\text{ㄱ}$, with fixed initial species ratios $\gamma = M/S$ and $\lambda = L/M$.

In this model, we introduce an additional consumption (grazing) matrix $G$, which indicates the relation between the producer $\alpha$ and the resource $\text{ㄱ}$.
$E_\text{ㄱ}$ indicates the degradation rate of the resource $\text{ㄱ}$.
Additionally, the biotic resources are also subject to intraspecific suppression, denoted by $d_\alpha$.

We obtain the relations that those three abundances should follow at the steady state likewise in Eq.~\eqref{eq:Supple-DMFT_results},
\begin{equation}
\label{eq:Supple-steady_Three}
\begin{aligned}
0 &=N \left( g_\text{eff} - h_\text{eff} N + z_N \sqrt{\sigma_g^2 + \sigma_h^2 N^2} \right),\\[6pt]
0 &= R \left( r_\text{eff} - d_\text{eff} R + z_R \sqrt{\sigma_R^2 + \sigma_d^2 R^2}\right),\\[6pt]
0 &= K - E_\text{eff} - \lambda^{-1}\sigma_G^2 \chi A^2 + z_A \sqrt{\sigma_K^2 + \sigma_A^2 A^2},
\end{aligned}
\end{equation}
where the mean effective degradation rate is written as $E_\text{eff} = \mu_E + \lambda^{-1} \mu_G \langle R \rangle$, and variance as $\sigma_A^2 =\sigma_E^2 + \lambda^{-1} \sigma_G^2 \langle R^2 \rangle$.
$\mu_G$ and $\sigma_G$ indicate mean and deviation of $G$, respectively.
$h_\text{eff}$, $g_\text{eff}$, and $\sigma_g$ are defined in Sec.~\ref{supsubsec:Supple-II_C}, but $r_\text{eff}$, $d_\text{eff}$, and $\sigma_R^2$ have an additional term from abiotic resource as $r_\text{eff} = \mu_r - \gamma^{-1} \mu_C \langle N \rangle + \mu_G \langle A \rangle$, $d_\text{eff} = d + \gamma^{-1}\sigma_C^2 \langle\nu\rangle + \sigma_G^2 \langle\theta\rangle$, and $\sigma_R^2 = \sigma_r^2 + \gamma^{-1}\sigma_C^2 \langle N^2\rangle + \sigma_G^2 \langle A^2 \rangle$, where $\langle\theta\rangle = - \left\langle \frac{\partial A}{\partial E} \right\rangle = \left\langle \frac{(\sigma_A^2 A^2 + \sigma_K^2)A}{\lambda^{-1}\sigma_A^2 \sigma_G^2 \langle\chi\rangle A^3 + (\sigma_A^2 K + 2\lambda^{-1} \sigma_K^2 \sigma_G^2 \langle\chi\rangle)A + \sigma_K^2 E_\text{eff}} \right\rangle$ is the response function of externally supplied abiotic resource.
The response function of the biotic resource $\chi$ is defined as $\chi=\langle \frac{\partial R}{\partial r} \rangle = \left\langle \frac{\sigma_R^2 + \sigma_d^2 R^2}{\sigma_d^2 r_\text{eff} R + d_\text{eff} \sigma_R^2} \right\rangle$.

From Eq.~\eqref{eq:Supple-steady_Three}, we obtain the abundance distributions for three-trophic level ecosystems in the same way described in Sec.~\ref{supsubsec:Supple-II_C} as follow,
\begin{equation}
\label{eq:Supple-abundnace_distributions_Three}
\begin{aligned}
P(N) &= \frac{1}{\sqrt{2\pi}} \left| \frac{\sigma^2_h g_\text{eff}  N  + \sigma^2_g h_\text{eff} }{(\sigma^2_h N^2 + \sigma^2_g)^{3/2}} \right| \exp \left[- \frac{(h_\text{eff} N - g_\text{eff})^2}{2 ( \sigma^2_h N^2 + \sigma^2_g)} \right],\quad \text{for}~N>0,\\[15pt]
P(R) &= \frac{1}{\sqrt{2\pi}}\left| \frac{\sigma_d^2 r_\text{eff} R + \sigma_R^2 d_\text{eff}}{\left( \sigma_d^2 R^2 + \sigma_R^2\right)^{3/2}} \right| \exp \left[ - \frac{\left( d_\text{eff} R - r_\text{eff} \right)^2}{2 \left( \sigma_d^2 R^2 + \sigma_R^2 \right)}\right],\quad \text{for}~R>0,\\[15pt]
P(A) &= \frac{1}{\sqrt{2\pi}} \frac{\lambda^{-1}\sigma_A^2 \sigma_G^2 \langle\chi\rangle A^3 + (\sigma_A^2 K + 2 \lambda^{-1} \sigma_K^2 \sigma_G^2 \langle\chi\rangle)A + \sigma_K^2 E_\text{eff}}{\left( \sigma_A^2 A^2 + \sigma_K^2 \right)^{3/2}} \exp\left[ - \frac{\left( \lambda^{-1} \sigma_G^2 \langle\chi\rangle A^2 + E_\text{eff}A - K\right)^2}{2(\sigma_A^2 A^2 + \sigma_K^2)} \right],\quad \text{for}~A \geq 0.
\end{aligned}
\end{equation}
Note that abundance distributions of two consumers $P(N)$ and $P(R)$ show the same functional form.

In the case of $\sigma_h=\sigma_d=0$, response functions of consumer species are given by $\langle\nu\rangle = \phi_S / h_\text{eff}$ and $\langle\chi\rangle = \phi_M / d_\text{eff}$.
Consequently, the following two relations hold: $\phi_S = h\langle\nu\rangle + \sigma_C^2\langle\nu\rangle\langle\chi\rangle$ and $\phi_M = d\langle\chi\rangle + \gamma^{-1}\sigma_C^2\langle\nu\rangle\langle\chi\rangle + \sigma_G^2 \langle\chi\rangle \langle\theta\rangle$.
In the same way before in Eq.~\eqref{eq:Supple-limit_half}, we derive the inequality $0\leq \sigma_G^2 \langle\chi\rangle \langle\theta\rangle \leq \lambda/2$.
Based on these relations, we can further obtain the inequality: 
\begin{equation}
d\langle\chi\rangle - \gamma^{-1}h\langle\nu\rangle \leq \phi_M - \gamma^{-1}\phi_S < d\langle\chi\rangle - \gamma^{-1}h\langle\nu\rangle + 1/2.
\end{equation}

\end{CJK*}

%%%%%%%%%%%%%%%%%%%%%%
\bibliographystyle{apsrev4-2}

%%%%%%%%%%%%%%%%%%%%%%
\end{document}